\documentclass
{article}
\usepackage[utf8]{inputenc}
\usepackage{lineno}
\usepackage{amsmath}
\usepackage{amsfonts}
\usepackage[a4paper, twoside, left = 2.5 cm, right = 2.5 cm, top = 30mm, footskip=3cm]{geometry}
\usepackage{setspace}
\usepackage{imakeidx}
\usepackage{enumitem}
\usepackage{tikz}
\usepackage{multirow}
\usepackage{algorithm}
\usetikzlibrary{calc,patterns,angles,quotes}
\usepackage{graphicx}
\usepackage{fancyhdr}
\usepackage{algpseudocode}
\usepackage{hhline}
\usepackage{subcaption}
\usepackage{pgfplots} 
\usepackage{amsthm}
\usepackage{hyperref}
\usepackage[section]{placeins} 
\usepackage{makecell}
\usepackage[
backend=biber,
style=numeric,
sorting=none
]{biblatex}
\fancypagestyle{myplain}
{
  \fancyhf{}

  \cfoot{\thepage}
}

\makeatletter

\usepackage{hyperref}

\newtheorem{theorem}{Theorem}[section]

\newtheorem*{claim*}{Claim}

\newtheorem{definition}[theorem]{Definition}

\newtheorem{lemma}[theorem]{Lemma}

\newtheorem{problem}[theorem]{Problem}

\newtheorem{remark}[theorem]{Remark}

\renewcommand\paragraph{\@startsection{paragraph}{4}{\z@}%
                                     {-3.25ex\@plus -1ex \@minus -.2ex}%
                                     {1.5ex \@plus .2ex}%
                                     {\normalfont\normalsize\bfseries}}
\makeatother

\title{
Algorithms for Detecting Shape Defects in Concentrated Solar Power Plant Components
}

\author{%
  J.M.\ Díaz-Báñez\textsuperscript{*,†}\thanks{%
    \textsuperscript{*}Department of Applied Mathematics II,
    University of Seville, Spain.
    \texttt{\small \{dbanez, mirmednav\}@us.es}\\
    \textsuperscript{†}Corresponding author.}
  \and
  M.\ Medrán-Navarro\textsuperscript{*}
}
\date{}

\begin{document}





\maketitle

\begin{abstract}
Concentrated solar power (CSP) technology offers a dispatchable renewable energy solution, capable of bridging the gap between intermittent generation and grid demand through high-capacity thermal energy storage. However, the overall thermodynamic efficiency and cost-effectiveness of CSP plants depend crucially on the optical quality and precise geometric alignment of thousands of reflective surfaces spread across vast solar fields. Shape defect detection algorithms have become essential tools for automating mirror profilometry, tracking the flow spot and diagnosing alignment issues. This paper presents a novel framework for geometric shape detection techniques tailored to CSP applications. It introduces the notion of \emph{parabolicity}, which motivates a class of optimisation problems in this setting and suggests a possible direction for further work in computational geometry.
Both exact and approximate algorithms are proposed, together with an experimental study on synthetic datasets generated via simulation.
\end{abstract}

\section{Introduction}


As part of ongoing efforts to decarbonise the European Union's electricity and heating systems \cite{RocaReina2025}, Concentrated Solar Power (CSP) has attracted renewed interest as a renewable energy technology. Unlike photovoltaic systems, which convert sunlight directly into electricity, CSP uses concentrated solar radiation to generate heat. This heat can be stored and used to produce electricity at a later time, offering a degree of flexibility that can complement variable renewable sources such as photovoltaic and wind power.

\begin{figure}[thb]
\centering
\begin{subfigure}{.52\textwidth}
\includegraphics[width=0.8\linewidth]{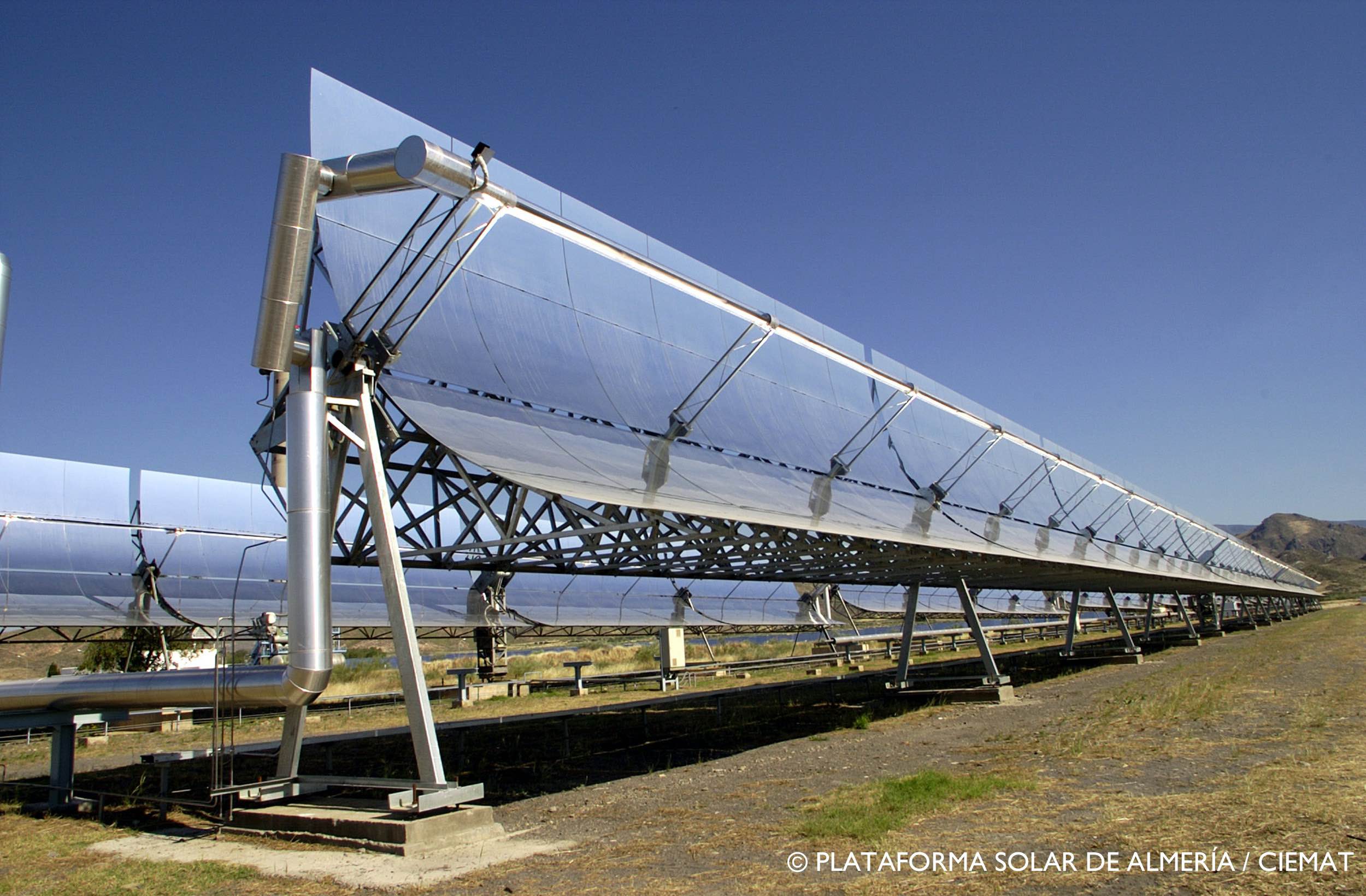}
\caption{Picture of a CSP \cite{PSAPhotoArchive}.}
\end{subfigure}%
\begin{subfigure}{.40\textwidth}
\includegraphics[scale =.14]{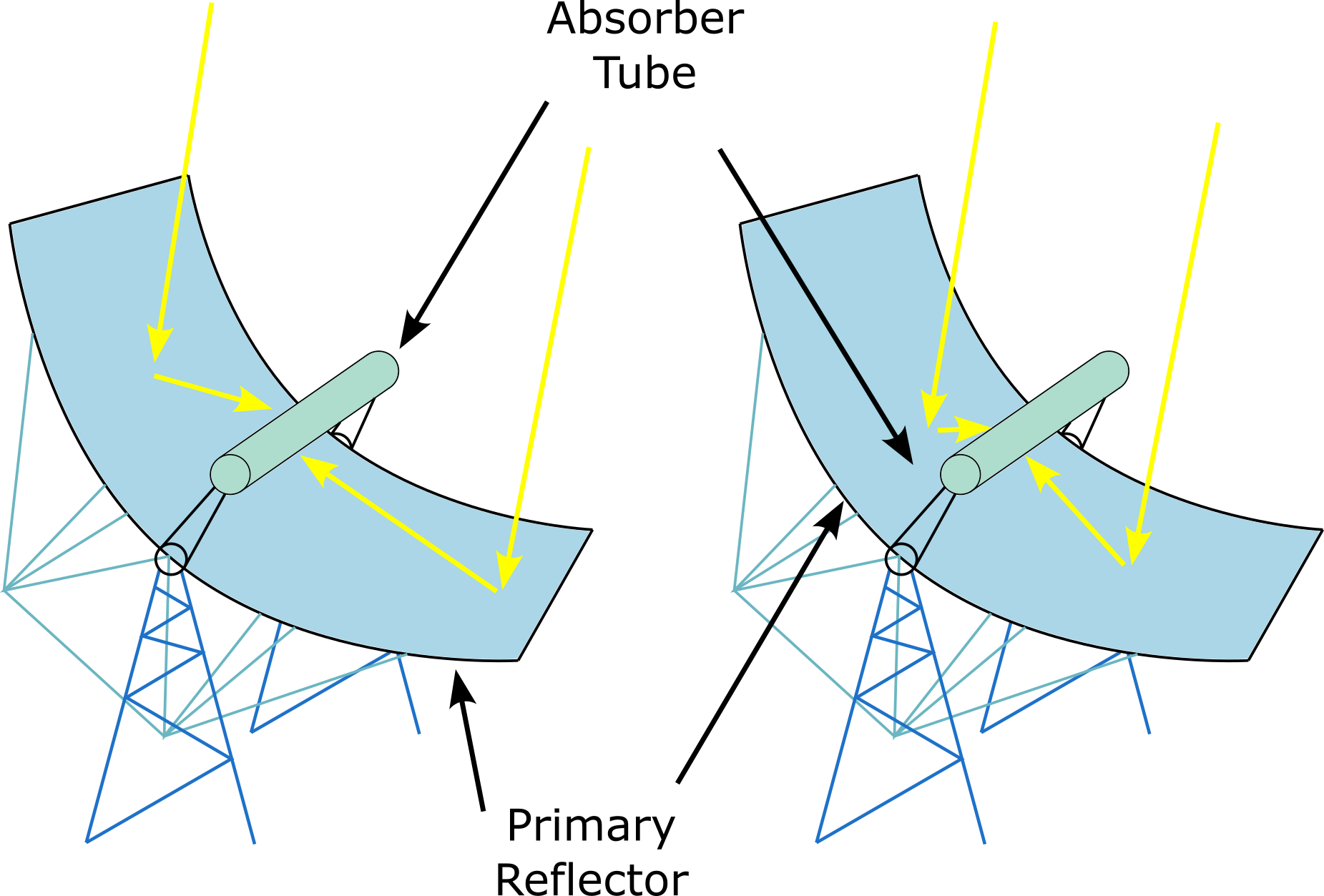}
\caption{Tube and reflector.}
\end{subfigure}%
\caption{Parabolic trough collector.}
\label{fig:intro}
\end{figure}

CSP plants employ different types of reflective collectors, including parabolic troughs and central receiver systems. In parabolic trough collectors (PTCs), curved mirrors focus incoming solar radiation onto a linear receiver tube positioned along the focal line of the reflector. The heat absorbed by the receiver is transferred to a working fluid and subsequently used for power generation \cite{LecuonaNeumann2023}. A typical parabolic trough collector is shown in Figure~\ref{fig:intro}.

\begin{figure}[h]
    \centering
    \begin{tikzpicture}

\def\f{3}      
\def\r{0.5}    

\draw[->, dashed] (-5,0) -- (5,0) node[right] {$x$};
\draw[->, dashed] (0,-0.5) -- (0,4) node[above] {$y$};

\draw[ultra thick,domain=-5:5,samples=200]
    plot (\x,{(\x)^2/(4*\f)});

\fill (0,\f) circle (2pt);
\node[xshift=-5pt,yshift=5pt] at (0,\f) {$F$};

\draw[ultra thick] (0,\f) circle (\r);

\coordinate (P) at (5,{25/(4*\f)});




\coordinate (tim) at (0, 0);
\coordinate (pim) at (-5,{25/(4*\f)});
\coordinate (Q) at (0,{\f)});
\draw[dashed] (pim) -- (Q);
\pic [draw, thick, <->, "$\theta_r$", angle radius=0.9cm, angle eccentricity = 1.25] {angle = pim--Q--tim};


\fill (tim) circle (2pt);
\node[below right] at (tim) {$V$};

\end{tikzpicture}
    \caption{Cross-section of a parabolic trough reflector, showing the receiver
    positioned at the geometric focus of the parabolic profile.}
    \label{fig:sectionDiagram}
\end{figure}
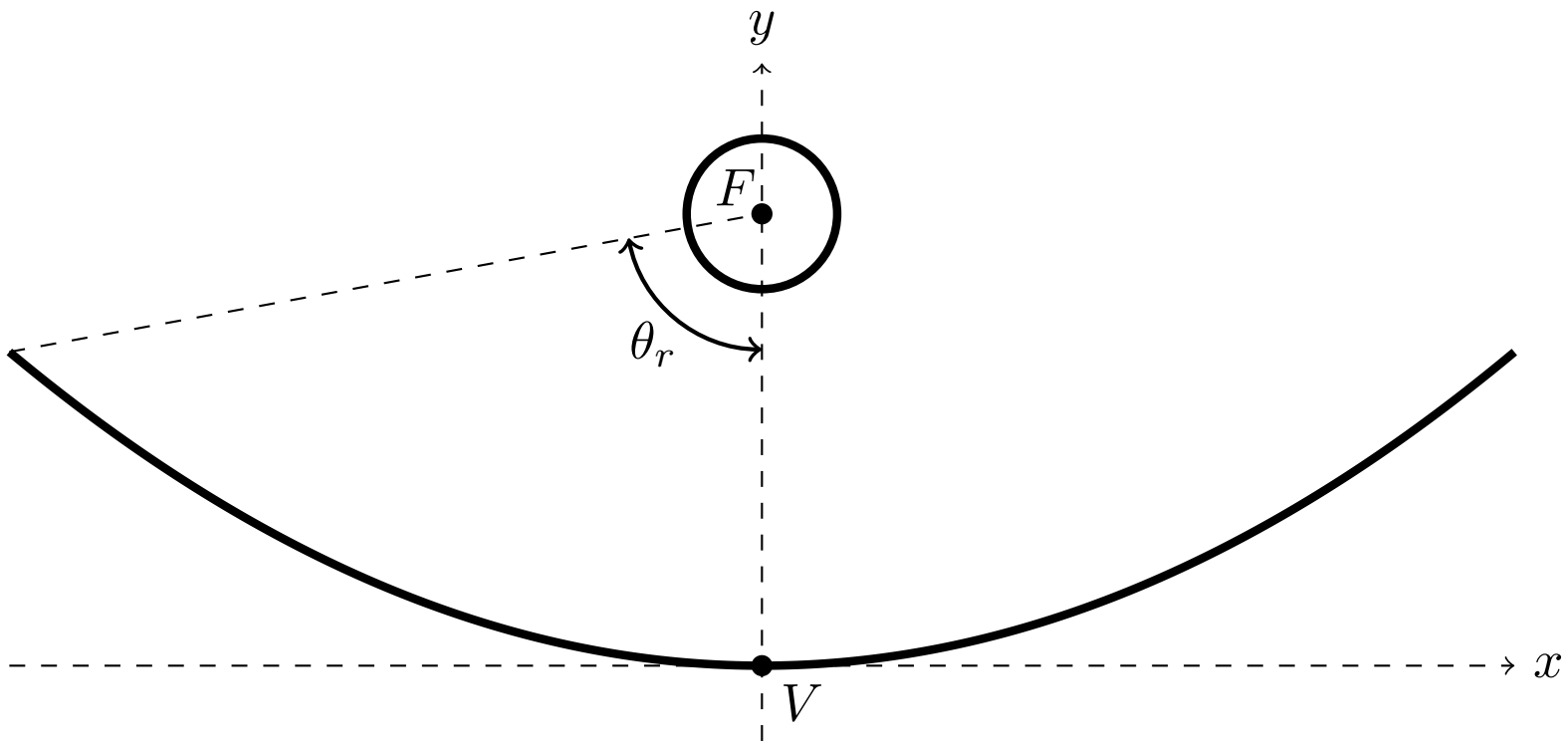

The operating principle of a parabolic trough collector relies on the focusing property of a parabola. Due to the large distance between the Sun and the Earth, incoming solar radiation can be approximated as parallel rays. When these rays are parallel to the parabola's axis of symmetry, they are reflected towards its focus. In a parabolic trough, this focusing property extends along the length of the reflector, concentrating the radiation onto the receiver tube (Figure~\ref{fig:sectionDiagram}).

The optical performance of parabolic trough collectors depends on the accuracy of their reflector geometry. Deviations from the nominal parabolic shape can alter the direction of reflected rays and reduce the amount of radiation intercepted by the receiver. Such deviations may arise from manufacturing imperfections, gravitational sagging, or structural deformation during operation. Their geometric characterization is therefore relevant to assessing the condition of the reflector.

In this work, reflector geometry is characterized using point clouds derived from laser scanning.
The primary objective of this work is to develop and evaluate geometric shape detection algorithms, specifically formulated for concentrating solar power environments. The scope of this work spans theoretical geometric optimization and experimentation. We introduce the concept of \emph{parabolicity} as a geometric measure of deviation from an ideal parabolic shape. Different computational algorithms are developed and compared to estimate this quantity, which can subsequently be used to define a form tolerance for the reflector. The proposed approaches are evaluated with particular attention to their ability to identify simulated geometric defects.

The main contributions of this paper are:
\begin{enumerate}
    \item \emph{Conceptual and Algorithmic Formalization of Parabolicity:} We introduce a mathematical framework for computing exact and approximate parabolicity metrics for 2D curves and 3D surfaces under noisy field observations.
    \item \emph{Algorithms Design:} We propose both exact and approximation approaches for automatic feature extraction, boundary detection, and surface slope reconstruction of solar facets. Efficient optimization methods for measuring parabolicity in discrete point sets, bridging the gap between abstract computational geometry and practical solar engineering.
    
    \item \emph{Validation and Performance Analysis:} We study the performance of the proposed algorithms on synthetic datasets.
\end{enumerate}


The remainder of this paper is organized as follows:
Section 2 reviews existing optical measurement techniques in CSP, standard geometric shape detection literature, and classic conic section fitting routines in computational geometry.
Section 3 defines the parabolic annulus and
the parabolicity of a point set, argues for the notion of width we adopt,
and states the problem in three and two dimensions.
Section 4 presents two different algorithms to solve the problem in two dimensions when the axis of symmetry is given.
Section 5 includes an exact algorithm for the non-oriented problem in two dimensions, as well as two algorithmic approximations.
Section 6 establishes the  experimental setup, explaining the steps followed to generate point clouds of parabolic troughs, the injected errors and the measure of their parabolicity.
Section 7 shows the findings of the experiments, and discusses its limitations.
Finally, Section 8 summarizes the core findings and outlines open avenues for future research in computational geometry for renewable energy.

\section{Related work}

Classic measures of mirror defects in PCTs include identifying slope errors through optical tests; however, alternative methods have been studied in the last few decades to identify geometric deviations and potential degradation issues associated with them. This includes applications such as photogrammetry \cite{photogrammetry}, which is a technique used to study and define an object's dimensions and shape through one or multiple images; and deflectometry, used to inspect specular surfaces by studying the reflections they produce of certain objects. An emergent case study is the use of laser scanning technologies to produce 3D models of the parabolic trough structures, whose geometry can be later studied. For instance, the work presented by De Asís López et al \cite{deAsisLopez2014} shows that terrestrial laser-scanning devices can be used to approximate the geometry of the mirror structure when measured from the back of the trough, that is, its non-reflective side, as the specular surface of the mirror, as seen from the front, can distort the image. In the work of Salamanca et al \cite{tsl2018} we see another use of laser scans of a parabolic trough, where the intercept and other geometric deviations are evaluated by comparing the point cloud to a \textit{Canonical Image} composed by the ideal manufacturing parameters of the trough structure. 
Guidi et al \cite{guidi} also provided a method involving laser-scanning the rear part of a trough to compare it against a Finite Element model and measure the deformations present, evaluating the accuracy of the process.

More recent works include \cite{tsl2025}, where two different laser scanning devices are used to obtain point clouds of the same solar power plant. The advanced software of the devices used corrected mirror deformations in the laser measurements; and part of their results include comparing two different scans, performed 18 months apart, to defect any differences in between. The main defects identified were systematic deformations at the upper and lower edges of the parabolic profile.

In all of these articles, the point clouds of the parabolic trough are obtained with a precision of the order of a few millimeters, i.e. Guidi et al \cite{guidi} show an uncertainty of less than 3.6 mm, and the setups for the laser scanning devices are stated to be simpler than those of alternatives such as photogrammetry or deflectometry. Therefore, identification of defects in mirrors through aerial or terrestrial laser scanning methods proves to be an interesting alternative, along with the need to devise formulations to measure and categorize the errors observed in the resulting point clouds.

In this work, we introduce a novel concept to measure the error defect in parabolic cylinder shapes. This measure, as we define in the following section, is inspired by existing notions such as cylindricity, known in metrology as a unit of tolerance for cylindrical shapes \cite{metrologyzheng}, \cite{har2001approximate}; and the minimum-width annulus, which is similarly used for determining tolerance in annular ring applications. The current ISO 12180-1:2011 already defines different terms surrounding the concept of cylindricity, including the measure of local cylindricity deviation as the deviation of a point on a cylindrical surface to the reference cylinder, with said deviation being normal to the reference cylinder \cite{ISOcilindricity}.

\begin{figure}[!htbp]
    \centering
    \begin{tikzpicture}


\draw [->] (0,-3.5) -- (0,1);
\node[xshift= 7, yshift= -5pt] at (0,1) {\footnotesize $y$};

\draw [dashed] (0,0) ellipse (1.25 and 0.5); 
\draw [dashed] (0,0) ellipse (0.75*1.25 and 0.65*0.5);
\draw [dashed] (0,-3.5) ellipse (1.25 and 0.5);
\draw [dashed] (0,-3.5) ellipse (0.75*1.25 and 0.65*0.5);
    
\draw [dashed] (-1.25,0) -- (-1.25,-3.5); 
\draw [dashed] (1.25,-3.5) -- (1.25,0);   
\draw [dashed] (-1.25*0.75,0) -- (-1.25*0.75,-3.5); 
\draw [dashed] (1.25*0.75,-3.5) -- (1.25*0.75,0);   
\draw [dashed] (1.25,-3.5) -- (1.25,0); 

\fill [gray!70,opacity=0.5,even odd rule] (0,0) ellipse (1.25 and 0.5) (0,0)
ellipse (0.75*1.25 and 0.65*0.5); 
\fill [gray!70,opacity=0.5,even odd rule] (0,-3.5) ellipse (1.25 and 0.5) (0,-3.5) ellipse (0.75*1.25 and 0.65*0.5); 

\fill [gray!70,opacity=0.5] (-1.25,0) -- (-1.25,-3.5) arc (180:360:1.25 and -0.5) -- (1.25,0) arc (0:180:1.25 and -0.5);

\draw[thick, smooth] plot coordinates {
(-1.1,-0.012)
(1.141 -2.2,-0.35)
(1.188 -2.2,-0.7)
(0.993 -2.2,-1.05)
(1.17 -2.2,-1.4)
(1.035 -2.2,-1.75)
(1.06 -2.2,-2.1)
(1.146 -2.2,-2.45)
(1.065 -2.2,-2.8)
(1.039 -2.2,-3.15)
(-1.1,-3.502)
};

\draw[thick, smooth] plot coordinates {
(1.117,0)
(1.241,-0.35)
(1.188,-0.7)
(0.993,-1.05)
(1.17,-1.4)
(1.035,-1.75)
(1.06,-2.1)
(1.146,-2.45)
(1.065,-2.8)
(1.039,-3.15)
(1.1,-3.503)
};

\draw[thick, smooth] plot coordinates {
(-1.1,-0.012)
(-0.99,0.169)
(-0.88,0.234)
(-0.77,0.266)
(-0.66,0.316)
(-0.55,0.334)
(-0.44,0.365)
(-0.33,0.381)
(-0.22,0.391)
(-0.11,0.392)
(0,0.389)
(0.11,0.38)
(0.22,0.381)
(0.33,0.365)
(0.44,0.347)
(0.55,0.344)
(0.66,0.3)
(0.77,0.284)
(0.88,0.227)
(0.99,0.173)
(1.1,-0.006)
(1.035,-0.187)
(0.92,-0.267)
(0.805,-0.302)
(0.69,-0.34)
(0.575,-0.37)
(0.46,-0.395)
(0.345,-0.402)
(0.23,-0.431)
(0.115,-0.427)
(0,-0.433)
(-0.115,-0.433)
(-0.23,-0.423)
(-0.345,-0.405)
(-0.46,-0.4)
(-0.575,-0.371)
(-0.69,-0.354)
(-0.805,-0.302)
(-0.92,-0.263)
(-1.035,-0.201)
(-1.1,-0.012)
};

\draw[thick, smooth] plot coordinates {
(-1.1,-3.502)
(-0.99,-3.686)
(-0.88,-3.75)
(-0.77,-3.8)
(-0.66,-3.822)
(-0.55,-3.86)
(-0.44,-3.868)
(-0.33,-3.898)
(-0.22,-3.895)
(-0.11,-3.915)
(0,-3.909)
(0.11,-3.911)
(0.22,-3.903)
(0.33,-3.882)
(0.44,-3.883)
(0.55,-3.86)
(0.66,-3.822)
(0.77,-3.796)
(0.88,-3.755)
(0.99,-3.686)
(1.1,-3.503)
(1.035,-3.333)
(0.92,-3.263)
(0.805,-3.215)
(0.69,-3.17)
(0.575,-3.155)
(0.46,-3.123)
(0.345,-3.104)
(0.23,-3.108)
(0.115,-3.095)
(0,-3.095)
(-0.115,-3.093)
(-0.23,-3.102)
(-0.345,-3.116)
(-0.46,-3.133)
(-0.575,-3.148)
(-0.69,-3.17)
(-0.805,-3.2)
(-0.92,-3.26)
(-1.035,-3.324)
(-1.1,-3.502)
};

\end{tikzpicture}
    \caption{Depiction of cylindricity deviation, confining the object between two concentric cylinders.}
\end{figure}
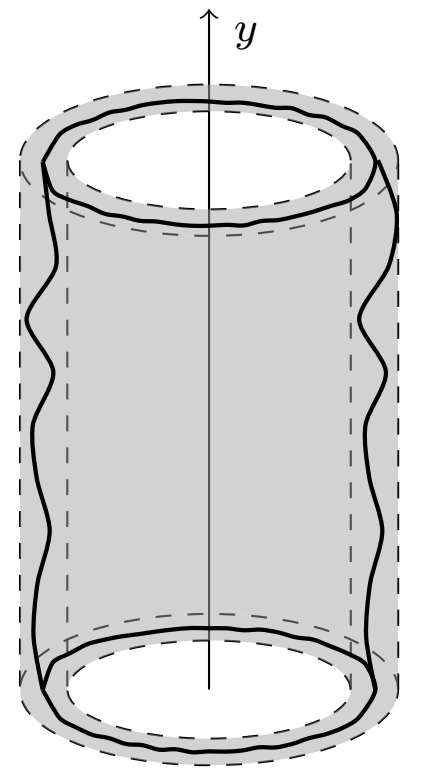

As mentioned earlier, cylindricity, as a definition, is also a way to measure roundness of a piece and as such, it finds its place in industries where high-level precision is required, for example, in fitting rotary parts \cite{rotary}. Definitions of cylindricity such as the Minimum-Zone Cylinder rely on the geometric properties of the cylinder, instead of being statistical averages. They define tolerance zones within a certain geometric feature, and compared to fittings performed with least-squares methods, the latter tend to overestimate form tolerance, possibly resulting in unnecessary rejections \cite{ZHANG2011484}.

The computation of these measures and regions is not trivial and has been an object of study for decades. In particular, extensive research has been conducted on finding linear and convex formulations to solve these problems, effectively reducing computational complexity. When properly defined, geometric problems can reduce their original polynomial complexity. One of the most well-known results in this field of study is that fixed-dimension linear problems can be solved in linear time in the number of constraints $n$ \cite{agarwal2001exact}, and it is widely applied in $\varepsilon$-approximate solutions as efficient alternatives \cite{har2001approximate}. 

Another well-known method in solving minimum-zone problems defines them as Chebyshev best-fit algorithms, whose solutions for the most common shapes---stripes, disks, spheres, cylinders---are characterized by the contacting points to the boundaries of the  solution. For a finite set of points, these problems become combinatorial in nature, and existing optimization software based on the Kuhn-Tucker equations was already capable of calculating approximations of them, as exposed in \cite{chebyshevbestfit}. Other computational geometry techniques also include Voronoi diagrams-based approaches \cite{voronoi}, constructing robus algorithms.

In the field of quadric surfaces, we remark on the work of Zhang et al \cite{ZHANGquadric}, which evaluates the minimum zone of form errors through a heuristic algorithm for multiple 3D quadrics, including parabolic cylinders. Har-Peled and Varadarajan \cite{harpeled} offer a general method to obtain linear $\varepsilon$-approximations of shape-fitting algorithms that can also be generalized for certain minimum zone problems. Lastly, although not presented explicitly as a measure of minimum zone errors, other works explore algorithms to fit discrete data points to parabolic shapes by minimizing specific parabolic features, such as \cite{LOPEZRUBIO2018301}.

Still, the fitting of parabolic and, more generally, quadric surfaces remains an open problem. There exist multiple ways to define geometric tolerances of the said shapes and each one constitutes a separate best-fitting problem to be studied on its own.

\section{Problem statement}

\subsection{Preliminaries}
In general, a parabola in a plane is defined as the locus of points equidistant from a fixed point, called the \textit{focus}, and a line, called the \textit{directrix}.

In a system of coordinates $XY$, the vertical parabola has the classic form:
\begin{equation*}
    y = ax^2+bx+c \quad \text{where } a\neq0
\end{equation*}

Whereas, when no orientation is given, it is represented by the general conic equation:
\begin{equation*}
    Ax^2+Bxy+Cy^2+Dx+Ey+F=0 \quad \text{where } B^2-4AC=0
\end{equation*}

A parabolic cylinder is a three-dimensional surface obtained by extending a planar parabola along a fixed direction parallel to the plane's normal. Equivalently, it is generated by translating the parabola along a direction perpendicular to its plane---the \textit{extrusion} axis---so that the cross-section perpendicular to this direction remains unchanged.

For example, a parabolic cylinder whose cross-section is $y=ax^2+bx+c$ and whose extrusion direction is the $z$-axis can be written as
\begin{equation*}
y=ax^2+bx+c,\qquad (x,y,z)\in\mathbb R^3.
\end{equation*}

\begin{definition}
A parabolic annulus is defined as the locus of points in 3D located between two parabolic cylinders. The width of a parabolic annulus is the separation distance between the two curved surfaces.
\end{definition}

The first notion that comes to mind is least squares, but it comes with certain caveats. A least-squares fit returns the parabola minimizing the sum of the squared residuals, which describes the data on average. On the other hand, a form tolerance asks a different question, namely how narrow a zone must be in order to contain every measured point. The annulus obtained by taking the extreme residuals of a least-squares fit does contain the points, so it is at least as wide as the minimum. For this reason, it tends to overstate the form error, and in a quality control setting that leads to rejecting parts which are in fact within tolerance \cite{ZHANG2011484}. The minimum zone is the definition that matches the question, and it is the one we adopt here.

\begin{figure}[h]
\centering
\begin{subfigure}{.5\textwidth}
  \centering
  \begin{tikzpicture}
\begin{axis}[
    axis lines=left,
    xmin=-2.3, xmax=2.3,
    ymin=-0.3, ymax=4.4,
    width=8cm,
    height=5cm,
    axis line style={black!70},
    ticks=none,
    clip=false,
]

\addplot[
    domain=-2:2,
    samples=150,
    thick,
    black,
] {x^2};

\addplot[
    domain=-1.67:1.67,
    samples=150,
    thick,
    black,
] {x^2+1.2};

\addplot[
    domain=-0.15:4.7,
    samples=2,
    densely dashed,
    gray,
    thick,
] ({0},{x});

\addplot[
    only marks,
    mark=*,
    mark size=1.6pt,
    black,
] coordinates {
    (-1.75,3.40)
    (-1.5,3.40)
    (-1.25,2.09)
    (-1,1.85)
    (-0.75,0.57)
    (-0.5,1.44)
    (-0.25,0.57)
    (0.25,1.18)
    (0.5,0.3)
    (0.75,1.31)
    (1,1.16)
    (1.25,2.22)
    (1.5,2.80)
};

\draw[<->, very thick]
    (axis cs:0,0) --
    (axis cs:0,1.2)
    node[midway,right=3pt] {$h$};

\end{axis}
\end{tikzpicture}
  \caption{Annulus based on vertical offset}
  \label{fig:measure1}
\end{subfigure}%
\begin{subfigure}{.5\textwidth}
  \centering
  \begin{tikzpicture}
\begin{axis}[
    axis lines=left,
    xmin=-2.3, xmax=2.3,
    ymin=-0.3, ymax=2,
    width=8cm,
    height=5cm,
    axis line style={black!70},
    ticks=none,
    clip=false,
]

\addplot[
    domain=-2:2,
    samples=150,
    thick,
    black,
] {0.25*x^2};

\addplot[
    domain=-1.6:1.6,
    samples=150,
    thick,
    black,
    black,
] {0.5*x^2 + 0.5 };

\addplot[
    domain=-0.3:2,
    samples=2,
    densely dashed,
    gray,
    thick,
] ({0},{x});

\addplot[
    only marks,
    mark=*,
    mark size=2.5pt,
    black,
] coordinates {(0,1)};

\node[anchor=west] at (axis cs:0.08,1)
    {$F$};
    
\addplot[
    only marks,
    mark=*,
    mark size=1.6pt,
    black,
] coordinates {
(-1.75,1.32)
(-1.5,1.6)
(-1.45,1.2)
(-1.35,0.49)
(-1.25,0.71)
(-1,0.4)
(-0.75,0.29)
(-0.5,0.25)
(-0.25,0.53)
(0.25,0.09)
(0.5,0.5)
(0.75,0.17)
(0.87,0.7)
(1,0.93)
(1.25,0.95)
(1.5,0.67)
(1.75,1.52)
};

 \draw[<->, very thick]
  (axis cs:0,0) --
   (axis cs:0,0.5)
   node[midway,right=3pt] {$\Delta f$};

\end{axis}
\end{tikzpicture}
  \caption{Annulus based on a shared focus}
  \label{fig:measure2}
\end{subfigure}
\makebox[\textwidth][c]{%
  \begin{subfigure}[t]{.48\textwidth}
    \centering
    \begin{tikzpicture}
\begin{axis}[
    axis lines=left,
    xmin=-2.1, xmax=2.1,
    ymin=-0.8, ymax=3,
    width=8cm,
    height=5cm,
    axis line style={black!70},
    ticks=none,
    clip=false,
]

\addplot[
    domain=-1.6:1.6,
    samples=200,
    thick,
    black,
] {x^2};
\node at (axis cs:1.85,2.25) {$p(x)$};

\addplot[
    domain=-1.2:1.2,
    samples=200,
    thick,
    dashed,
    black,
] (
    {x - 0.6*x/sqrt(1+4*x^2)},
    {x^2 + 0.3/sqrt(1+4*x^2)}
);

\addplot[
    domain=-1.2:1.2,
    samples=200,
    thick,
    dashed,
    black,
] (
    {x + 0.6*x/sqrt(1+4*x^2)},
    {x^2 - 0.3/sqrt(1+4*x^2)}
);

\draw[<->, thick]
    (axis cs:0,0) --
    (axis cs:0,0.3)
    node[pos=0.74,right=3pt] {$\delta$};

\end{axis}
\end{tikzpicture}
    \caption{Annulus based on normal distance}
    \label{fig:measure3}
\end{subfigure}%
}
\caption{Sectional view of different width examples}
\label{fig:test}
\end{figure}
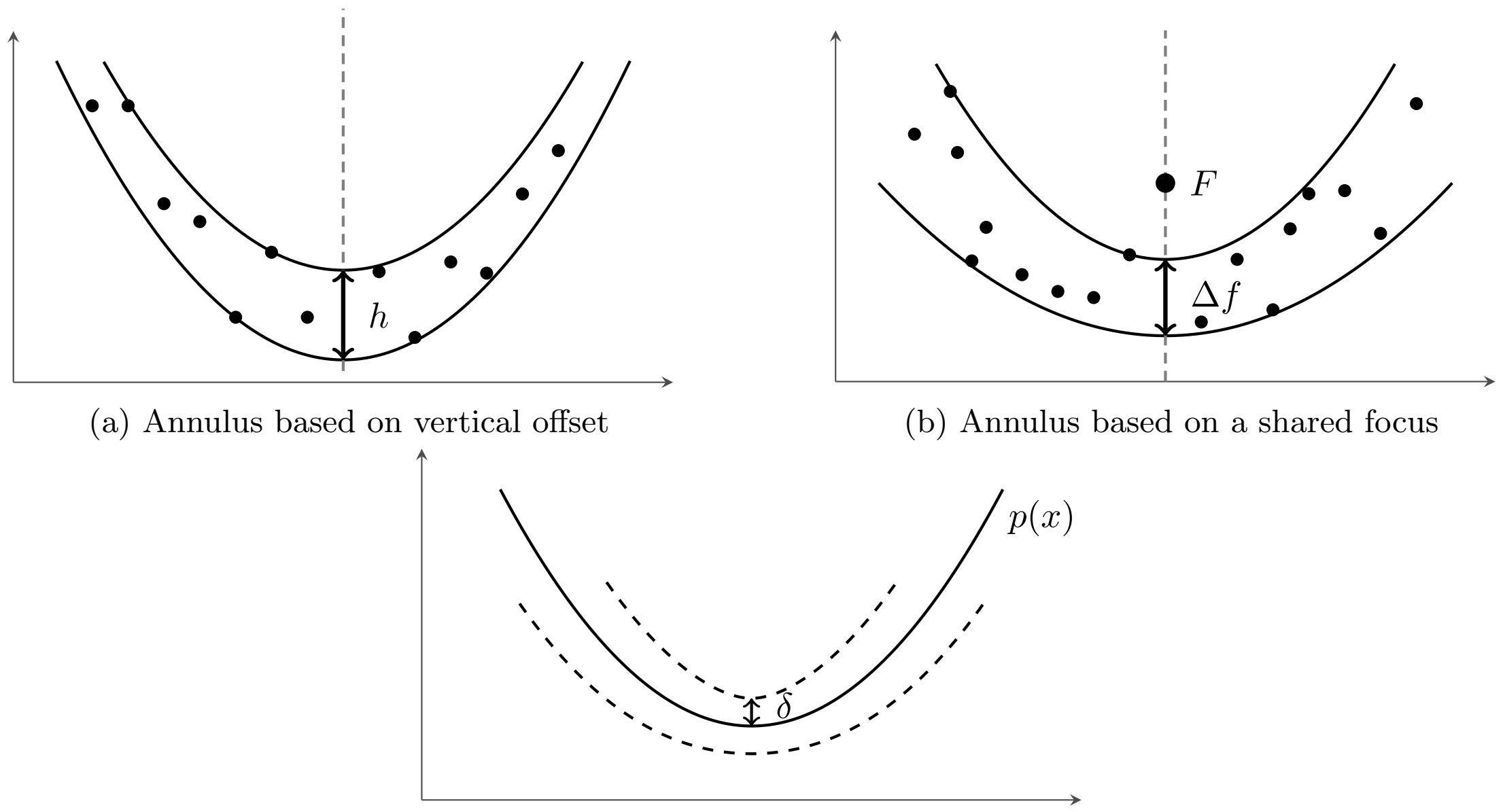

There are many possible definitions of width, the definition depending on the
metric considered. For example, if both parabolic cylinders share the same
axis of symmetry and one surface is a shift of the other, the width is the
distance along the axis direction. Another possibility is to fix the focus and
take the width as the variation in focal distance between the bounding
parabolas. We illustrate some examples in Figure \ref{fig:test}.

Note that if we select the width of our annulus as the normal distance from a reference or central parabola---respectively, in 3D, a parabolic cylinder---as in orthogonal-distance fitting, the resulting region is no longer, in general, a parabolic annulus. Assuming that the enclosing surfaces are cylindrical extensions of planar curves, their planar cross-sections are \textit{offset} curves of the reference parabola. By Corollary 2.3 of Farouki and Neff \cite{FAROUKI1990101}, the offset curve of a non-degenerate parabola has algebraic degree six. Consequently, the resulting enclosing surfaces are not, in general, parabolic cylinders.

In this work, we consider the width as the distance along the axis direction and explore different methods to compute it. Taking two identical parabolas displaced along their common axis leads to simpler formulations, because the deviation of a point from the parabola can then be linear. We exploit this in the sections that follow.

With the following definition, we introduce the main object of study of this work.

\begin{definition}
We say that a set of points $S$ has parabolicity $\rho$ if the width of the minimum parabolic annulus containing $S$ is equal to $\rho$.
\end{definition}

\subsection{Problem statement in 3D}

Although there are multiple ways to define parabolicity as a minimum-zone measure, this thesis focuses on a specific construction based on two identical parabolas translated relative to one another along their axis of symmetry. In three dimensions, this construction extends to two parabolic cylinders, yielding the following formal problem.

\begin{problem}
    Given a finite set of points $S\subseteq\mathbb{R}^3$, find a minimum-width parabolic annulus containing $S$, bounded by two identical parabolic cylinders translated relative to one another along the axis of symmetry of their planar profiles. The width of the annulus is defined as the magnitude of this translation, and the minimum width is the parabolicity of $S$.
\end{problem}

\subsection{Problem statement in 2D}

Assuming the extrusion axis of the parabolic cylinder is known, we can project our set of points $S\subseteq \mathbb{R}^3$ onto a plane orthogonal to this axis. This reduces the original problem to a 2D instance, which can then be solved using planar techniques. It will be the main scope of this work.

\begin{figure}[htbp]
\centering
\begin{tikzpicture}[>=stealth, scale=1.15]

    \fill[blue!8]
        plot[domain=0.3:5.3, smooth, variable=\x] ({\x},{0.2*(\x-2.5)^2+1})
        -- plot[domain=5.3:0.3, smooth, variable=\x] ({\x},{0.2*(\x-2.5)^2+3})
        -- cycle;

    \draw[->, thin] (-0.3,0) -- (6.1,0);
    \draw[->, thin] (0,-0.3) -- (0,5.0);

    \draw[dashed, gray] (2.5,0) -- (2.5,4.25)
        node[above, black, font=\small] {axis of symmetry};

    \draw[domain=0.3:5.3, smooth, variable=\x, blue!70!black, thick]
        plot ({\x},{0.2*(\x-2.5)^2+1});
    \draw[domain=0.3:5.3, smooth, variable=\x, red!70!black, thick]
        plot ({\x},{0.2*(\x-2.5)^2+3});
    \node[right, blue!70!black] at (5.35,2.568) {$p(x)-t/2$};
    \node[right, red!70!black]  at (5.35,4.568) {$p(x)+t/2$};

    \draw[<->, thick] (1.5,1.2) -- (1.5,3.2);
    \node[left, font=\small] at (1.5,2.2) {$t$};

    \foreach \px/\py in {%
        0.60/2.822, 0.95/1.880, 1.30/2.888, 1.65/1.845, 2.00/2.350,
        2.35/1.304, 2.70/2.808, 3.05/1.861, 3.40/2.562, 3.75/1.813,
        4.10/3.212, 4.45/2.761, 4.80/2.658, 5.15/3.605}
        \filldraw[black] (\px,\py) circle (1.7pt);

    \foreach \px/\py in {1.00/1.450, 3.50/1.200}
        \filldraw[fill=white, draw=blue!70!black, thick] (\px,\py) circle (2.4pt);
    \foreach \px/\py in {2.00/3.050, 4.60/3.882}
        \filldraw[fill=white, draw=red!70!black, thick] (\px,\py) circle (2.4pt);

\end{tikzpicture}
\caption{The minimum-width parabolic corridor of a point set. The two
boundaries are translates of one another along the axis of symmetry, and $t$ is
the width to be minimised. The open circles mark the points attaining the
boundaries.}
\label{fig:corridor}
\end{figure}
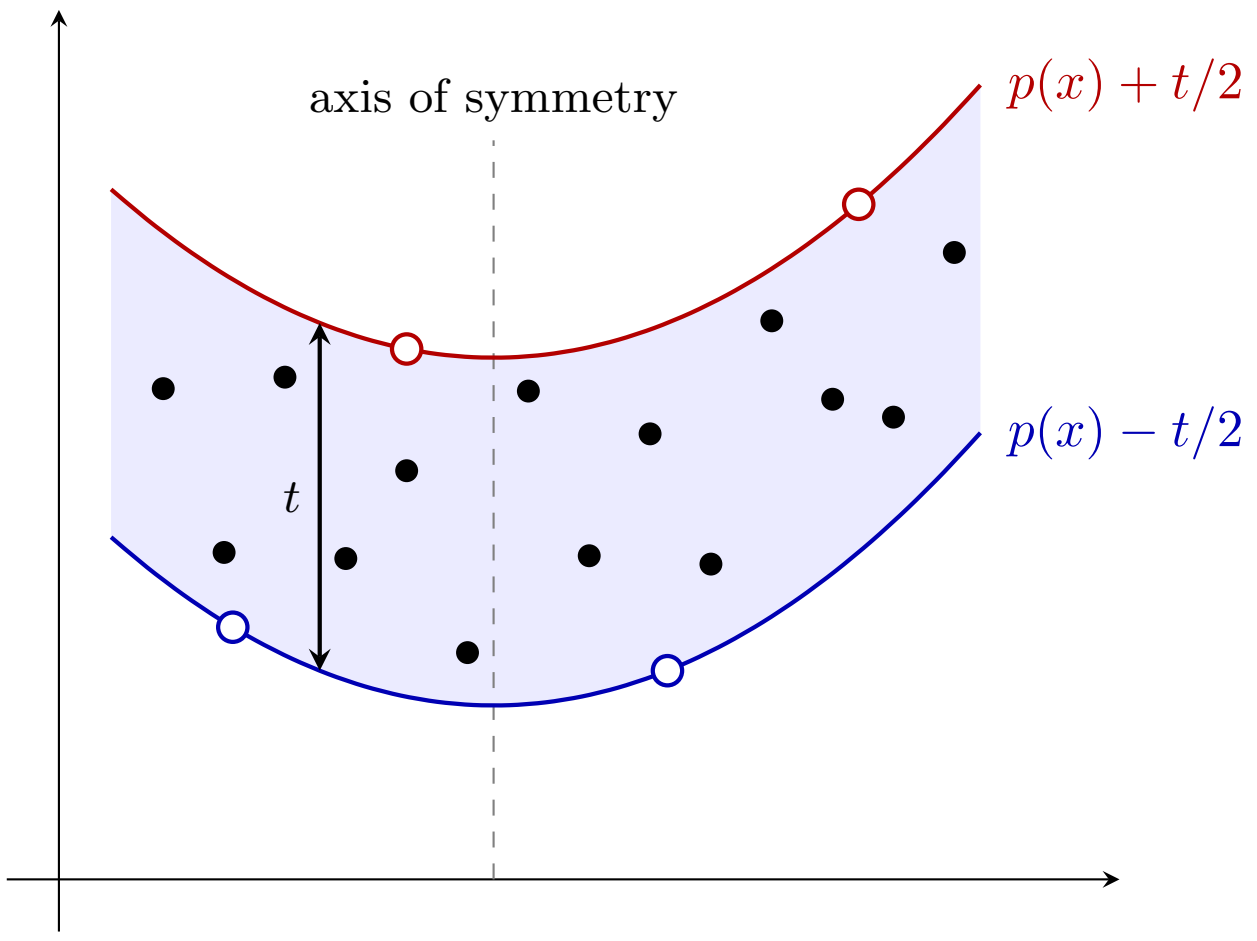

\begin{definition}
A parabolic corridor in 2D is defined as the locus of points in the plane located between two parabolas. The width of a parabolic corridor is the separation distance between the two curves measured vertically along the axis of symmetry. See Figure \ref{fig:corridor}.
\end{definition}

\begin{problem}\label{prob:2d} (Minimum-Width Parabolic Corridor with Variable Axis Direction).
Given a set $P = \{p_1, p_2, \ldots, p_n\}$ of $n$ points in $\mathbb{R}^2$, where each point $p_i = (x_i, y_i)$, compute a direction vector ${d} = (\cos \theta, \sin \theta)$ and two parabolas sharing a common axis given by said direction, such that the width of the parabolic corridor is minimized.

\end{problem}
\begin{remark}
We distinguish between two edge cases and a third one where the point set is already assumed to have a vaguely parabolic form, depending on the aperture $a$ of the parabolas.
\begin{enumerate}
    \item When $a \to 0$: The parabolas become straight lines, reducing to the problem of finding the minimum-width strip containing all points.
   \item When $a \to \infty$: The parabolas become highly curved, and the optimal location of the vertex will tend to place vertices near the points.
   \item For points roughly following a parabolic shape but with some noise, the optimal orientation will align the parabola axis with the natural axis of the point distribution, minimizing the required thickness of the parabolic annulus.
\end{enumerate}
\end{remark}

\newpage
\section{Algorithms for the oriented problem}

The first version is to consider that the axis direction is known. Without loss of generality, we assume the axis of the parabolas is vertical (you can rotate the set of points later if the axis is oblique). Thus, The two parabolas will have the form:
\begin{itemize}
    \item Lower parabola: $y_u=ax^2+bx+c_1$
    \item Upper parabola: $y_l=ax^2+bx+c_2$,
\end{itemize}

where $a$ and $b$ are the same for both, since one is a translation of the other. The objective is to minimize the vertical difference $t=c_2-c_1$, under the condition that for each point $(x_k, y_k)$ in the set:
$$ax_k^2+bx_k+c_1\leq y_k\leq ax_k^2+bx_k+c_2.$$

\subsection{A linear programming method}

This problem can be efficiently solved using Linear Programming, since all constraints are linear with respect to the parameters we want to find $(a, b, c_1, t)$.

With all the considerations above, we have the following formulation:

\begin{equation} \label{eq:lp1}
    \begin{split}
    &\mathrm{min} \quad c^Tx  \\[2pt] 
    &\mathrm{s.t.} \quad 
    \begin{array}[t]{r c r}
         Ax & \leq & b\\
     \end{array}
     \end{split}
\end{equation}

Where $c = (0,0,0,1)^T$ and $x=(a,b,c_1,t)^T$, the constraints are defined as follows:

\begin{align*}
     &A_{i*} = \begin{cases}
     \begin{pmatrix}
            \hphantom{-}x_{i\hphantom{-n}}^2 &  \hphantom{-}x_{i\hphantom{-n}} & \;1 & \hphantom{-}0
        \end{pmatrix} & \text {for } i = 1,\dots,n \\
       \begin{pmatrix}
         -x_{i-n}^2 & -x_{i-n} & \;-1 & -1
        \end{pmatrix}  & \text{for } i =n+1,\dots,2n
    \end{cases} \\[2pt]
     &b_{i*} = \begin{cases}
    \hphantom{-}y_i & \text{for } i = 1,\dots,n \\
    -y_{i-n} & \text{for } i =n+1,\dots,2n
    \end{cases} \\[4pt]
     &\text{with } A \in \mathbb{R}^{2n \times 4},\; b \in \mathbb{R}^{2n \times 1} 
\end{align*}

That is, we have $2n$ constraints. The first $n$ constraints determine the lower bound of our parabolic annulus and the following $n$ constraints define the upper bound of the annulus. 

Note that it is not necessary to impose $t\geq=0$ since it is obtained from adding the corresponding inequalities given in the restriction $Ax\leq b$ for every $i$ and $i+n$ rows. 

\begin{theorem}\label{th:lp}
The oriented version of the Problem \ref{prob:2d} can be solved in $O(n)$ time.
\end{theorem}
\begin{proof}
To prove feasibility, consider any finite set of points $\{(x_k,y_k)\}_{k}$. Then we can set $a=b=0$, $c_1=\min_k{y_k}$ and $t = \max_k{y_k} - \min_k{y_k}$, which always satisfies the condition $Ax\leq b$. 
Since our objective function is bounded below and we are minimizing over a closed set $P=\{x\in\mathbb{R}^4: Ax\leq b\}$, the optimum is attainable and finite.

We have now proved our $2n$ boundaries in (\ref{eq:lp1}) define a feasible set in $\mathbb{R}^4$. Additionally, as the dimension $d$ is fixed ($d=4$), Megiddo's result \cite{agarwal2001exact} can be applied to solve this in $O(n)$ time. 
\end{proof}

\subsection{An alternative algorithm}

The following result gives a useful characterization of the support point set. It is inspired by Chebyshev polynomial approximations and their equioscillating properties \cite{cheney1966}.

\begin{lemma}
\label{lemma:equioscillation}
Given a set of points $\{(x_k,y_k)\}_{k=1}^n$ with distinct abscissae and the problem described in (\ref{eq:lp1}) with a unique annulus solution, there exist four support points, $x_{k_1}<x_{k_2}<x_{k_3}<x_{k_4}$ , ordered by abscissa, lying alternately on the upper and lower boundaries of the minimizing annulus.
\end{lemma}

\begin{proof}
Let $p(x)$ be the central parabola that describes the shape of the minimizing annulus, with $t$ being the minimum parabolicity and width of said annulus, and let $S$ the set of support points. Since our program has a fixed dimension of $d=4$, we know, by the properties of linear programs, that there ought to be at least four points in our set $\{(x_k,y_k)\}_{k=1}^n$ where the equality holds exactly:
\begin{equation*}
    Ax=b,
\end{equation*}
which is the same as saying that for these points, $p(x_k) = y_k \pm t/2$.
By contradiction, let us assume that in our support group there are at most two pairs of consecutive points where the alternating property holds true. For these pairs of consecutive points, say $x_k$ and $x_{k+1}$, we pick a point $z_k \in (x_k,x_{k+1})$. At most, we will have two new points $z_k$, since our support set needs to contain at least one point on each boundary, both upper and lower. Then exists a second degree polynomial $q(x)$ such that $q(z_k)=0$ for all $z_k$. 

Now let us define $\tilde{p}(x) = p(x) + \varepsilon q(x)$ where $\varepsilon >0$ is small enough to still be inside the annulus. Take:
\begin{equation*}
    \varepsilon = \frac{1}{2} \min_k \{|y_k - p(x_k)-\frac{t}{2}|,|y_k - p(x_k)+\frac{t}{2}|\} \quad \forall x_k \notin S
\end{equation*}
It follows that $\varepsilon \neq 0$. We also can scale $q(x)$ over our point set so that $1=\max_k{|q(x_k)|}$.
Then it holds that $|\tilde{p}(x) - y_k| < t/2$ in $x_k \notin S$. We can impose on our $q(x)$ that $\text{sign}(q(x_k)) = \text{sign}(y_k-p(x_k))$ is true for all points $x_k \in S$, by how its roots were selected earlier.
\begin{enumerate}
    \item If $y_k - p(x_k) = t/2$ then $\tilde{p}(x_k) = p(x_k) + \varepsilon q(x_k) > p(x_k) = y_k -t/2 $ and $|\tilde{p}(x_k) - y_k| < t/2$.
    \item If $y_k - p(x_k) = -t/2$ then $\tilde{p}(x_k) = p(x_k) + \varepsilon q(x_k) < p(x_k) $ and $|\tilde{p}(x_k) - y_k| < t/2$.
\end{enumerate}
Therefore, the new polynomial $\tilde{p}(x)$, which defines a parabola because it is a polynomial of degree 2, yields a parabolicity of less than $t$, which is a contradiction, and the support points are alternating. 
\end{proof}

Note that this proposition confirms the existence of the said four support points, but there may be other points that intersect with the boundaries of the annulus. This proposition only asserts that we can find, among said points, a subset with the alternating property: upper, lower, upper, lower, or vice versa.

The following result can be used to reduce the number of constraints in the case where $a$ is positive, that is, when the optimal parabola is convex.

\begin{lemma}
\label{lem:lowerSupport}
Let $\mathcal{P}=\{(x_k,y_k)\}_{k=1}^{n}$ be a finite point set and let $p$ be a
convex function with
\begin{equation*}
    y_k \;\ge\; p(x_k), \qquad k=1,\dots,n .
\end{equation*}
If the constraint is active at the index $j$, that is $y_j=p(x_j)$, then
$(x_j,y_j)$ lies on the convex hull of $\mathcal{P}$. If $p$ is strictly
convex, $(x_j,y_j)$ is a vertex of that hull.
\end{lemma}

\begin{proof}
Suppose $(x_j,y_j)$ does not lie on the convex hull of $\mathcal{P}$, and let
$(x_i,y_i)$, $(x_l,y_l)$ be the endpoints of the edge of the lower hull
spanning $x_j$, so that $x_j=\lambda x_i+(1-\lambda)x_l$ for some
$\lambda\in(0,1)$ and $\lambda y_i+(1-\lambda)y_l<y_j$, since otherwise $(x_j,y_j)$ would be on the convex hull. Then, by the
convexity of $p$ and the feasibility of those two points,
\begin{equation*}
    p(x_j)\;\le\;\lambda\,p(x_i)+(1-\lambda)\,p(x_l)
          \;\le\;\lambda y_i+(1-\lambda)y_l
          \;<\;y_j ,
\end{equation*}
contradicting $y_j=p(x_j)$, so an active point must be a vertex.
\end{proof}

Whereas the two previous lemmas provided properties of the points that defined the optimal solutions, the following gives a sufficient condition for optimality. 

\begin{lemma}\label{lemma:valle} Given a set of points $\{(x_k,y_k)\}_{k=1}^n$ and let $p(x)=ax^2+bx+c$ be a second-degree polynomial. Define $e = \max_{k=1,\dots,n}|y_k - p(x_k)|$, and assume there are four points $(x_{k_l},y_{k_l})$ with $l=1,\dots,4$ ordered such that $x_{k_1}<x_{k_2}<x_{k_3}<x_{k_4}$ and that:
\begin{enumerate}[label=(\roman*), leftmargin=1.5cm]
    \item $e = |y_{k_l} - p(x_{k_l})|$ for $l=1,\dots,4$.
    \item $\text{sg}\left(y_{k_l} - p(x_{k_l})\right) = - \text{sg}\left(y_{k_{l+1}} - p(x_{k_{l+1}})\right)$, that is, the residual amounts alternate in sign. 
\end{enumerate}
Then $p(x)$ is an optimal solution for (\ref{eq:lp1}).
\end{lemma}

\begin{proof}
    To prove this lemma, we first note that the minimum attained by problem (\ref{eq:lp1}) for the given point set is equivalent to writing:
    \begin{equation}\label{eq:minmax1}
        t = 2 \min_{p \in \Pi_2} \max_{1\leq k \leq n} |y_k - p(x_k)|
    \end{equation}
    Where $\Pi_2=\{p(x)=ax^2+bx+c \ | \ a,b,c\in \mathbb{R} \}$ is the set of polynomials of degree at most two.
    
    Now, let $\rho$ be the parabolicity associated to the problem (\ref{eq:lp1}) and $\hat{p}$ is the central parabola of the minimum width-annulus of our point set. We will show that
    \begin{equation*}
        \rho \geq 2 e = 2 |y_{k_l} - p(x_{k_l})|
    \end{equation*}
    By contradiction, assume that this is not true and $\max_k |y_k - \hat{p}(x_k)| < e$. Then for each $(x_{k_l},y_{k_l})$ it follows that:
    \begin{equation*}
        p(x_{k_j}) - \hat{p} (x_{k_j}) = (y_{k_j}-\hat{p} (x_{k_j})) - (y_{k_j}-p (x_{k_j}))
    \end{equation*}
    By (i), we have that $|y_{k_l}-\hat{p} (x_{k_j})|<e=|y_{k_j}-p (x_{k_j})|$, which means $\text{sg}(p(x_{k_j}) - \hat{p} (x_{k_j}))=-\text{sg}(y_{k_j}-p (x_{k_j}))$. Since these values alternate at $x_{k_1}<\dots<x_{k_4}$, by our assumption (ii), $p-\hat{p}$ also alternates sign at these four points. But since $p-\hat{p}$ is a second-degree polynomial that changes sign at three different points, it must have three distinct roots. Therefore $p-\hat{p} \equiv 0$, but $\max_k |y_k - \hat{p}(x_k)| < e$, which is absurd. Therefore $\rho \geq 2e$, while the polynomial $p$ of our hypothesis satisfies that $2e = 2\max_{k=1,\dots,n}|y_k - p(x_k)|$, meaning that $p$ is optimal for (\ref{eq:minmax1}) and returns the minimum-width annulus. 
\end{proof}

Lemmas \ref{lemma:equioscillation} and \ref{lemma:valle} characterize optimality, and both lead to a result that makes it possible to implement Remez's exchange algorithm \cite{remez1934}. This algorithm is built on iteratively solving the exchange system for four ordered points, selected at a time, and computing the maximum error for the parabola that the exchange system returns. For this to work, the exchange system has to return a unique parabola, which the following Lemma proves.

\begin{lemma}
\label{lemma:exchange}
Let $x_{j_1}<x_{j_2}<x_{j_3}<x_{j_4}$ be four distinct
abscissae. Then the following system, known as the exchange system,
\begin{equation}\label{eq:remezSystem}
    a x_{j_i}^{2}+b x_{j_i}+c+(-1)^i h=y_{j_i}
    \qquad i=1,\ldots,4,
\end{equation}
has a unique solution $(a,b,c,h)\in\mathbb{R}^4$.
\end{lemma}

\begin{proof}
It suffices to show that the associated homogeneous system has only the trivial solution. Suppose
\begin{equation*}
    q(x_{j_i})=-(-1)^i h\qquad i=1,\ldots,4,
 \end{equation*}
where $q(x)=ax^2+bx+c$. If $h\neq0$, the values of $q$ at the four
ordered abscissae alternate in sign. By continuity, $q$ must have a root in each of the three disjoint intervals between consecutive
abscissae. This is impossible for a nonzero polynomial of degree at most two. Hence $h=0$. It follows that $q$ vanishes at four distinct points, so $q\equiv0$. Therefore $a=b=c=h=0$, and the system is non-singular.
\end{proof}

\subsubsection{Remez's exchange algorithm}

The Remez algorithm \cite{remez1934} is an iterative algorithm designed to solve minimax approximation problems. It constructs an approximation over a finite set of $n$ points and progressively builds improvements according to the error attained. To find a best-fitting polynomial of degree at most two, it can be described by the following steps:
\begin{enumerate}
    \item Select four ordered points $x_1<x_2<x_3<x_4$.
    \item Solve the exchange system described in Lemma \ref{lemma:exchange}. It will give a polynomial $p(x)$ of degree two or less and an error level $h$.
    \item Evaluate the errors over the entire point set $e_k=p(x_k)-y_k$., and calculate the maximum error, which will be denoted $E_i=\text{max}_{k=1,\dots,n}|e_k|$, with $i$ being the current iteration.
    \begin{enumerate}
        \item If $E_i$ equals $h$ up to a numerical tolerance, that is, if the maximal error already occurs at our four ordered points, then the algorithm stops, and we have found the optimal solution that satisfies the equioscillation property of Lemma \ref{lemma:equioscillation}.
        \item If $E_0 >h$, then we choose a new set of four ordered points where the errors alternate in sign. It should include the points with the largest errors. The algorithm is repeated with another iteration, solving the new exchange system. 
    \end{enumerate}
    \item The procedure is terminated when a predetermined stopping criterion is satisfied. For example, it may stop when the change in the maximum error between successive iterations satisfies $|E_i-E_{i-1}|<\tau_{\mathrm{tol}}$ or when a maximum number of iterations is reached.
\end{enumerate}

Compared to the original system described by Remez, our problem does not approximate a continuous function, so the classic convergence arguments do not directly establish convergence of our implementation, and we cannot guarantee that the exchange procedure will reach the optimum. However, we can contrast the solutions given by our implementation of Remez with those given by the linear programming formulation in the experimental section of this work.

\section{Algorithms for the non-oriented problem}

In this section, we study the general case, where the axis of symmetry is unknown. We propose an algorithm to solve the problem exactly and and  two approaches to get approximations. A theoretical note for the first approximation approach is also included.

\subsection{Exact algorithm}
\label{sec:exact_algorithm}

To solve the non-oriented minimum width parabolic corridor problem exactly, we leverage the arrangement of algebraic candidate curves.

Let us assume that the pair of parameters $(a, b)$ of the parabola is fixed. At any rotation angle $\theta$, the vertical width required to enclose all points $P = \{p_1, \dots, p_n\}$ using a corridor with quadratic term $a$ and linear term $b$ is given by:$$f_{a,b}(\theta) = \max_{p_i \in P} \left( y_i(\theta) - a x_i(\theta)^2 - b x_i(\theta) \right) - \min_{p_j \in P} \left( y_j(\theta) - a x_j(\theta)^2 - b x_j(\theta) \right)$$where $(x_i(\theta), y_i(\theta))$ are the rotated coordinates of point $p_i$:$$x_i(\theta) = x_i \cos\theta + y_i \sin\theta$$$$y_i(\theta) = -x_i \sin\theta + y_i \cos\theta$$For a fixed choice of $(a, b)$, $f_{a,b}(\theta)$ is a continuous curve as $\theta$ varies over $[0, \pi)$.The optimal width function $W^*(\theta)$ is obtained by finding the best possible $(a, b)$ at each angle $\theta$:$$W^*(\theta) = \inf_{(a, b)} f_{a,b}(\theta)$$Geometrically, drawing the continuous curves $f_{a,b}(\theta)$ for every valid pair $(a, b)$, $W^*(\theta)$ forms the lower envelope of this family of curves.

Now, for a fixed orientation $\theta$, a parabolic corridor is determined by 4 parameters: Parabola curvature $a$; Linear slope $b$; Vertical shift $c$; Corridor width $w$. Since there are 4 degrees of freedom, any subset of 4 points $S = \{p_i, p_j, p_k, p_l\}$ (partitioned into top/bottom curves) uniquely determines the corridor parameters $(a, b, c, w)$ for that angle $\theta$.
As $\theta$ sweeps continuously from $0$ to $\pi$, this 4-point set traces out a continuous 1D width curve $f_{a,b}(\theta)$, which we denote by $W_S(\theta)$.
Thus, the non-oriented minimum width function across all angles is given by the \emph{lower envelope} of these candidate curves:
\begin{equation}
W^*(\theta) = \min_{S \subset P, |S|=4} W_S(\theta)
\end{equation}

\begin{lemma}
\label{lem:envelope_complexity}
The lower envelope $W^*(\theta)$ of the $N = O(n^4)$ candidate width curves $\{W_S(\theta)\}$ has a combinatorial complexity of $O(n^4 \beta(n))$, where $\beta(n)
$
is an extremely slowly growing function 
\end{lemma}

\begin{proof}
Any pair of distinct 4-point width curves $W_{S_1}(\theta)$ and $W_{S_2}(\theta)$ intersect at angles $\theta^*$ where both 4-point subsets define identical corridor parameters. At such intersections, $S_1 \cup S_2$ forms a 5-point touch configuration. Because $W_{S_1}(\theta)$ and $W_{S_2}(\theta)$ are bounded algebraic curves of constant maximum degree, the number of real intersection roots on $[0, \pi)$ is bounded by a constant $s = O(1)$. Using Davenport--Schinzel sequence bounds for $(N, s)$-sequences, the lower envelope consists of at most $\lambda_s(N) = O(N \cdot \beta(N))$ connected algebraic arcs \cite{sharir1988davenport}. Since $N = O(n^4)$ in our problem,   the proof is done.
\end{proof}

The following algorithm calculates the parabolicity exactly:

\vspace{.25cm}

{\textbf{Algorithm for minimum width non-oriented parabolic corridor:} 


\begin{enumerate}
    \item \textbf{Candidate Curve Generation:}
    For each of the $O(n^4)$ 4-point subsets $S \subset P$ and each partition of $S$ into upper and lower boundary sets, formulate the 4 linear equations in variables $(a, b, c, w)$ and obtain $W_S(\theta)$.
    
    \item \textbf{Divide-and-Conquer Envelope Construction:}
    Construct the lower envelope $W^*(\theta) = \min_S W_S(\theta)$ over $\theta \in [0, \pi)$ using divide-and-conquer:
    \begin{enumerate}
        \item Partition the set of $O(n^4)$ curves into two equal halves.
        \item Recursively compute the lower envelope for each half.
        \item Merge the two lower envelopes by running a 1D sweep-line algorithm along $\theta$ to identify all $O(n^4 \beta(n))$ intersection points.
    \end{enumerate}

    \item \textbf{Critical Point Minimization:}
    Identify the overall minimum width $w^*$ on the constructed lower envelope $W^*(\theta)$ by evaluating:
    \begin{enumerate}
        \item \emph{Local Minima (Stationary 4-Point Touches):} Solve $\frac{d W_S(\theta)}{d\theta} = 0$ along each active arc segment.
        \item \emph{Breakpoints (5-Point Critical Touches):} Evaluate $W^*(\theta)$ at each intersection vertex between adjacent arc segments.
    \end{enumerate}

    \item \textbf{Output:}
    Return the global minimum width $W^* = \min_{\theta} W^*(\theta)$ and its corresponding exact orientation $\theta^*$.
\end{enumerate}

\begin{theorem}
Algorithm 1 computes the exact minimum width non-oriented parabolic corridor in $O(n^4 \log n)$ time.
\end{theorem}

\begin{proof}
By Lemma \ref{lemma:equioscillation} and LP vertex properties, the global minimum must occur either at a stationary point of a 4-point corridor or at a 5-point transition vertex. The lower envelope $W^*(\theta)$ contains all such candidates by construction. The divide-and-conquer merge step takes time proportional to the envelope complexity and the recurrence running-time formula is:
\begin{equation}
T(N) = 2 T(N/2) + O(N \beta(N)) \implies T(n)\in O(n^4 \log n)
\end{equation}
Checking validity across all $n$ points for the candidate minima requires $O(n)$ verification per arc, yielding 
a naive total verification cost of $O(n^5 \beta(n))$. To maintain the $O(n^4 \log n)$, state-of-the-art computational geometry algorithms avoid checking points one-by-one at the end. 
Instead, to avoid testing all $n$ points ($O(n)$ time) at every candidate step, we can  restrict the 4-point curves $W_S(\theta)$ to their valid angular domains. This requires the use of tools for dynamic computational geometry updating. This approach maintains global containment while preserving the $O(n^4\log n)$ overall runtime.
\end{proof}

In the practical application that inspired the minimum parabolic annulus problem, the dataset is very large, which makes the exact algorithm described above impracticable. In what follows, we propose two efficient approximation methods that may prove useful in practice.

\subsection{Approximation 1: The grid discretization algorithm}

A simple approach can be built based on the oriented case by rotating the frame of reference in which our point cloud is defined. 

\begin{enumerate}
\item Iterate over $m$ equally spaced angles $\varphi \in [0,\pi)$.

\item For each $\varphi$, rotate the points and solve the Linear Programming problem previously defined in (\ref{eq:lp1}), using the new data points instead: \[\begin{cases}
    x(\varphi) = x\cos{\varphi} - y\sin{\varphi} \\
    y(\varphi)= x\sin{\varphi} + y\cos{\varphi}
\end{cases}\]

\item Keep the angle that gives the minimum width. The corresponding parabolicity value will be the approximation to the problem.
\end{enumerate}
Even though the parabolicity calculation computed in step 2 is exact, the final measure still holds a margin of error over the real parabolicity, as we are approximating the axis of symmetry, making it a heuristic rather than a  approximation with guarantee.

It is worth remarking that this algorithm does not rely on the pivot point on which our rotations are built---in this case, we chose the origin of the frame of reference of our point cloud---since, for a given direction, the lineal problem for the oriented case defined in (\ref{eq:lp1}) is invariant to any translation.  Following Theorem \ref{th:lp}, this approach requires $O(m\cdot n)$ time.

\subsection{Approximation 2: A linear approximation algorithm}

The following approximation is based on the assumption that the points should follow a reference parabola and that we know, approximately, where the axis of symmetry lies. Based on this, we can assume that any rotation we perform on our system will therefore be small. Similar linearization approximations are present in works for cilindricity \cite{devillers2000evaluating}.

Let us consider again the same problem as before, assuming the ideal parabolic trough is modeled by:

\begin{equation}
    ax^2 + bx + c = y \quad \text{where } a,b,c \in \mathbb{R} \text{ are known}
\end{equation}

In fact, when $b=0$, then $a$ becomes the aperture of the parabola. We assume that these parameters are known and that the data are observed in a coordinate frame that differs from the true symmetry-aligned frame by a small rotation. Then there exists a perturbation angle $\delta \in \mathbb{R}$ by which we can rotate our points $(x,y)$ to be written in the ideal frame $(x',y')$, and it follows that:
\[\begin{cases}
    x(\delta) = x\cos{\delta} - y\sin{\delta} = x - \delta y + O(\delta^2)\\
    y(\delta)= x\sin{\delta} + y\cos{\delta} = \delta x + y + O(\delta^2)
\end{cases}\]

With $|\delta| \leq \varepsilon$. Assuming $\delta$ is sufficiently small that we can neglect the terms of order $O(\delta^2)$, our problem then becomes solvable through an LP under the constraints:
\begin{equation}
\begin{aligned}
   & ax_k^2 - 2\delta ax_ky_k  - y_k - \delta x_k  \leq c & \forall k = 1,\dots,N\\
   & y_k + \delta x_k  -  ax_k^2 + 2\delta ax_ky_k  \leq d & \forall k=1,\dots,N\\
   &-\varepsilon \leq \delta \leq \varepsilon \\
\end{aligned}
\end{equation}
where we wish to minimize $\rho = c + d$ as our annulus width, or parabolicity approximation.

We can also generalize this to include perturbations in the horizontal offset $b$ and the aperture $a$ of our model, such that
\begin{equation}
    \begin{cases}
        a \rightarrow a + \Delta a +O(\varepsilon^2) \\
        b \rightarrow b + \Delta b + O(\varepsilon^2)
    \end{cases}
\end{equation}
With all perturbations satisfying  $|\delta|\leq\varepsilon$, $|\Delta a|\leq C_a\varepsilon$ and $|\Delta b| \leq C_b \varepsilon$. Again, we neglect terms of order $O(\varepsilon^2)$, to arrive to a new set of linear constraints, with $c,d,\Delta a, \Delta b$ and $\delta$ as variables, returning the following minimization problem.

\begin{problem}\label{prob:linearised} (Linearized parabolicity with bounded perturbations).
Let $P=\{(x_k,y_k)\}_{k=1}^{N}$, let $a,b$ be the nominal coefficients, let
$\varepsilon>0$ be a trust radius and $C_a,C_b\ge0$ constants. Writing
\begin{equation}\label{eq:pertResidual}
    r_k \;=\; \bigl(y_k-ax_k^{2}-bx_k\bigr)
          \;+\; \delta\,u_k
          \;-\; \Delta a\,x_k^{2}
          \;-\; \Delta b\,x_k ,
    \qquad \text{where } u_k \;=\; x_k+2a\,x_ky_k+b\,y_k ,
\end{equation}
compute $(\delta,\Delta a,\Delta b,c,t)$ solving
\begin{equation}\label{eq:lpPert}
\begin{aligned}
    \min_{\delta,\Delta a,\Delta b,c,t}\quad & t \\
    \text{s.t.}\quad & c \;\le\; r_k \;\le\; c+t,
        && k=1,\dots,N, \\
    & |\delta|\leq \varepsilon, \quad
      |\Delta a|\leq C_a\varepsilon, \quad
      |\Delta b|\leq C_b\varepsilon, \quad t\ge0 .
\end{aligned}
\end{equation}
\end{problem}

Without the assumption that perturbations are of similar order and of negligible squared terms, the first $2n$ constraints given by $c\leq r_k $ and $r_k \leq c+t $ become non-linear equations that are not trivial to solve. 

So far, in our presented algorithms, we have only worked with oriented frames of reference or small perturbations of said case. In truth, our measure of parabolicity $\rho$ is quite sensitive to rotations. Indeed, if we model a modest data set delimited by perfectly-spaced points that conform to two identical parabolas, except for a vertical displacement, we find that the parabolicity varies greatly under axis rotations.

\begin{remark} The problem given by \ref{prob:linearised} is linear. This is given by Megiddo's result \cite{megiddo1984} and the fact that the problem has fixed dimension, $d=5$.
\end{remark}

\begin{remark}
The formulation described in this section here does not, in general, strictly correspond to fitting a parabolic annulus over the set of directions $[-\varepsilon,\varepsilon]$. More specifically, the term $2\delta a x_k y_k$ introduces a mixed quadratic term without the corresponding $y_k^2$ term, so the resulting expression does not preserve the exact structure of the intended parabolic model, except when the optimal value of the relevant parameter is exactly zero.
\end{remark}

The same observation holds for the method of \cite{devillers2000evaluating}, in which a family of quadrics approximating cylinders with nearly vertical axes is combined with a linearized optimization formulation. The original geometric problem is replaced by a more manageable mathematical formulation.

\subsection{A note on the grid discretization algorithm}

{\bf Grid Discretization Algorithm:}
Assume an exact  algorithm $\mathcal{A}_{\text{oriented}}(P, \theta)$ for the oriented problem. To approximate the non-oriented problem, we construct a uniform angular grid $\Theta = \{\theta_1, \theta_2, \dots, \theta_m\} \subset [0, \pi)$ with step size $\Delta\theta = \frac{\pi}{m}$, execute $\mathcal{A}_{\text{oriented}}(P, \theta_i)$ for each $\theta_i \in \Theta$, and return:
\begin{equation}
\tilde{W} = \min_{\theta_i \in \Theta} W^*(\theta_i)
\end{equation}

The question is whether this grid discretization approach constitutes a  $(1+\epsilon)$-approximation algorithm for the non-oriented case.
We establish a negative result.


\begin{theorem}
\label{thm:negative}
The discrete angle sampling algorithm does not guarantee a relative $(1+\epsilon)$-approximation for the non-oriented minimum parabolic corridor problem in general. That is, for any fixed $m \in \mathbb{N}$, there exist point sets $P$ such that $\frac{\tilde{W}}{W^*} > 1 + \epsilon$.
\end{theorem}

\begin{proof}
    
Suppose the points of $P$ lie almost perfectly along a parabola with axis orientation $\theta^* \in [0, \pi)$. The minimal corridor width is $W^* = W^*(\theta^*) \approx 0$. 
Since $\theta^*$ is a continuous real parameter, unless $\theta^* \in \Theta$ explicitly, any sampled orientation $\theta_i \in \Theta$ ($\theta_i \neq \theta^*, i=1,\cdots m$) will yield a strictly positive width $W^*(\theta_i) > 0$. Consequently:
\begin{equation}
\frac{\tilde{W}}{W^*} = \frac{\min_{\theta_i \in \Theta} W^*(\theta_i)}{W^*} > 1 + \epsilon
\end{equation}
Since a finite grid $\Theta$ has measure zero in $[0, \pi)$, $\theta^*$ cannot be guaranteed to belong to $\Theta$, yielding an unbounded approximation ratio.
\end{proof}

\section{Experiments}

In this section, we describe the experimental setup, the steps of which
Figure \ref{fig:pipeline} sets out in the order they are performed. We explain how the point
clouds are generated, how defects are injected into the collector, and how a
tolerance against which to judge the resulting widths is obtained. The implementation
algorithms for both the oriented and non-oriented cases is also described. What they report is the subject of Section \ref{sec:results}.

\begin{figure}[htb]
    \centering
    \includegraphics[width=0.7\linewidth]{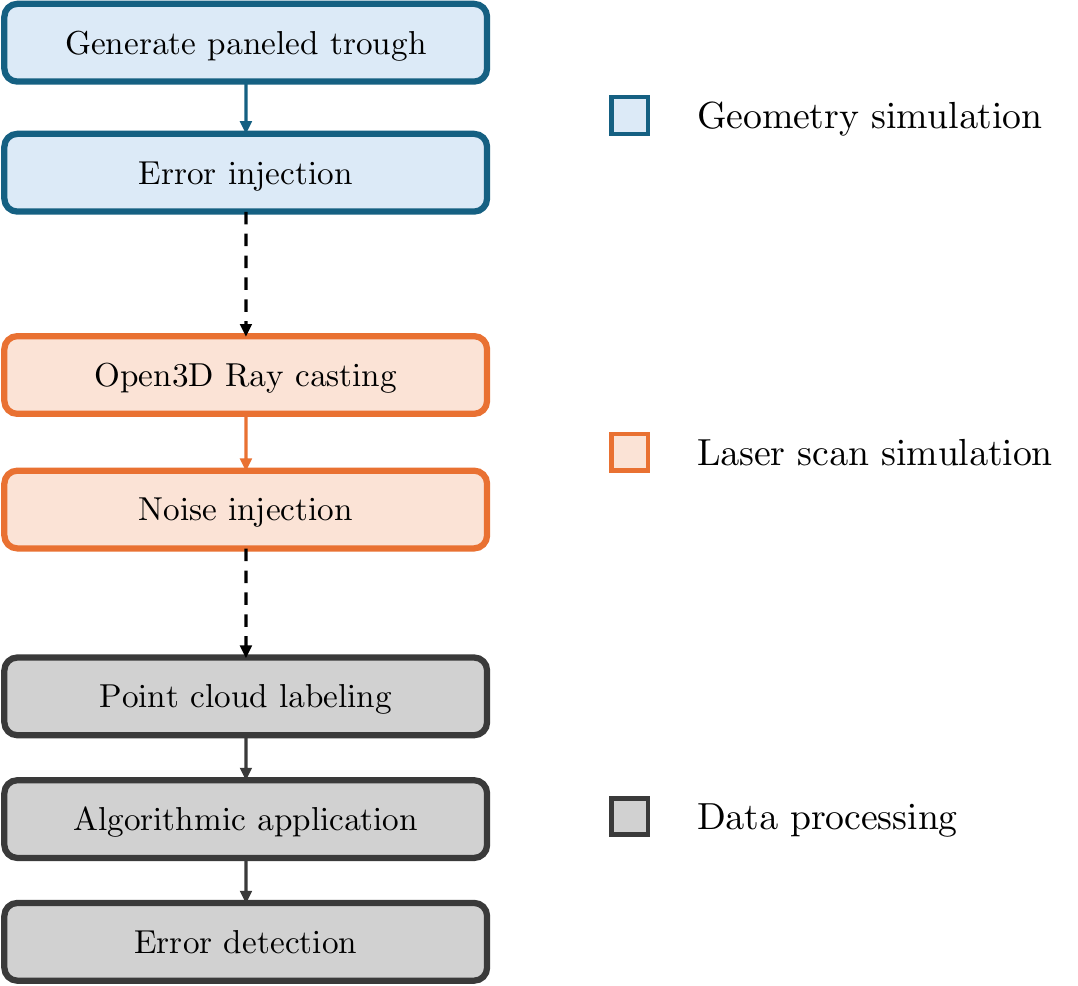}
    \caption{Diagram of the simulation pipeline.}
    \label{fig:pipeline}
\end{figure}

\subsection{Cloud points generation}

The point clouds used to evaluate parabolicity will be synthetic. Our scenario
is a laser scanner carried above a collector. Laser scanners have been applied
to the recovery of collector geometry both on prototypes
\cite{deAsisLopez2014, guidi} and on operating solar fields
\cite{tsl2018, tsl2025}. Having no measurements of a real collector at our
disposal, we simulate the scan instead. The arrangement we model is
deliberately generic, opting for a grid of viewpoints above the collector,
which describes a multi-station terrestrial survey as readily as an airborne
pass. Since the parabolicity is evaluated on the cross-section, any instrument
furnishing a dense sample of the profile would serve equally well.

Simulations of this kind are common in engineering, and proceed from a
three-di\-men\-sional model of the item to be measured, typically in a CAD format.
We use Open3D \cite{open3d}, an open-source library for C++ and Python for
processing three-dimensional data, whose ray-casting facilities reproduce a
scan from the number of pulses emitted and the trajectory of the device. The
sensor occupies $240$ positions---sixty stations along the collector at each of
four offsets across it, at a height of $3$ m above the vertex---and emits from
each a fan of $108^{\circ}$ across the collector and $20^{\circ}$ along it. Such path is known as a raster scan, and its diagram is depicted in Figure \ref{fig:raster}. It returns a dense point cloud across the profile.

\begin{figure}[htbp]
    \centering
    \includegraphics[width=0.7\linewidth]{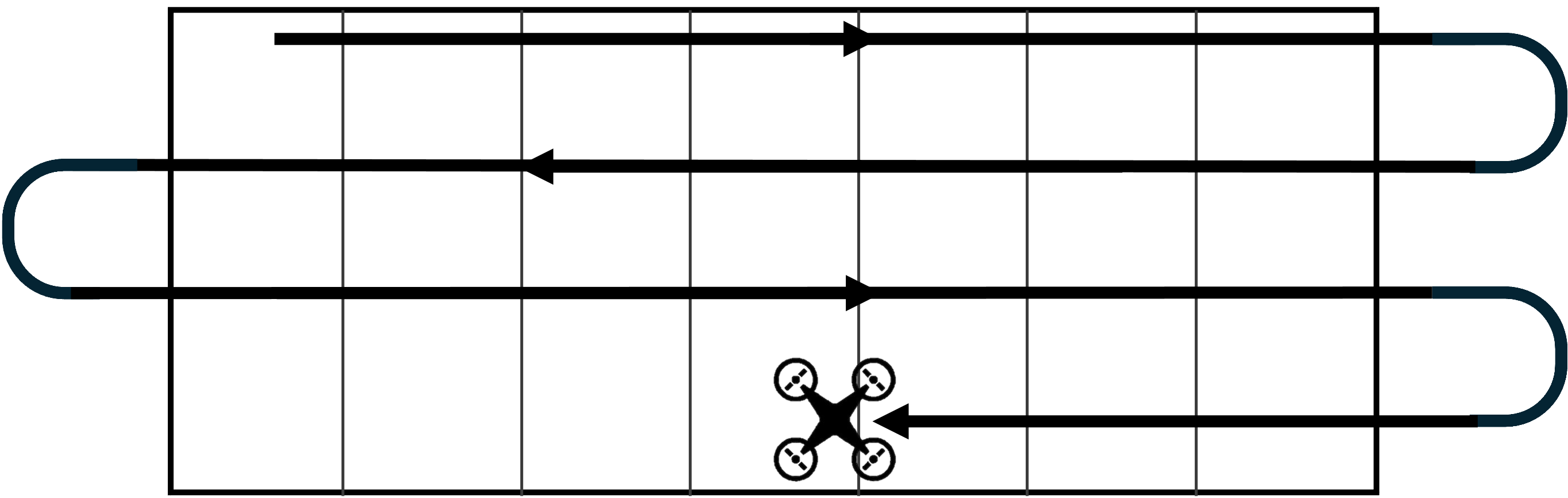}
    \caption{Schematic of raster scan. The drone flies along parallel lines over the field.}
    \label{fig:raster}
\end{figure}

Afterwards, random noise is injected into the resulting point cloud, using a Gaussian function to account for real-life inaccuracies. From now on, we consider this to be the data cloud output of our laser scan. Since laser scans provide information for the entire trough and its surroundings, labeling the points is crucial to isolate those belonging to the parabolic mirrors. However, these labels will be provided by the simulated data, as we know the original mesh structures. 

The next step is voxel down-sampling. The space is fractionated into identical cubes of a certain size, and for each cube (the \textit{voxel}), the centroid of the data points contained inside is computed. The resulting centroids will be the new cloud points. This step reduces possible duplicate points by using a voxel of $15$ mm.

Lastly, we project the points of the mirror onto a plane perpendicular to the
extrusion axis. This yields the planar cloud $\{(x_k,y_k)\}_k$ which is used as input for our algorithms and completes the pipeline of Figure \ref{fig:pipeline}.

To generate the paneled trough, we use the dimensions of the EuroTrough model \cite{eurotrough}, as presented in Table \ref{tab:eurotrough}.

\begin{table}[h]
\centering
\begin{tabular}{ |c|c| } 
 \hline
 Focal length & 1.71 m \\
 Aperture width & 5.76 m \\ 
 Module length & 12 m \\ 
 Number of panels across cylinder length & 7 \\ 
 Number of panels across cylinder width & 4 \\
 \hline
\end{tabular}
\caption{Main characteristic parameters used in the simulation.}
 \label{tab:eurotrough}
\end{table}

The baseline 3D model of the panel is presented in Figure \ref{fig:3dtrough}. Although not visible in the plot, the panels are modeled separately in order to identify and inject defects more easily.

\begin{figure}[htbp]
    \centering
    \includegraphics[width=0.55\linewidth]{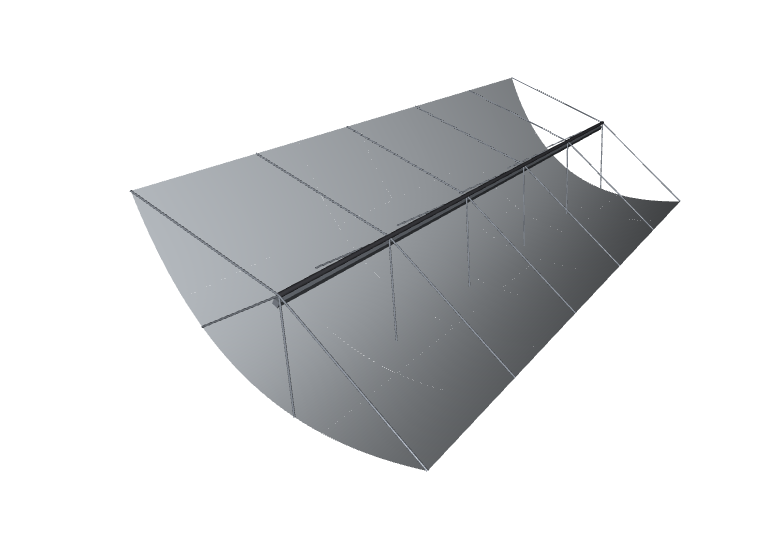}
    \caption{Simulated model with the EuroTrough dimensions.}
    \label{fig:3dtrough}
\end{figure} 

After running the simulation of the laser scan, we will obtain a point cloud that, when plotted in 3D, mostly conserves the parabolic shape of the trough, barring some possible occlusions that will be seen as empty spaces in the plot.

\newpage

\subsection{Simulation of errors}

Defects and errors in parabolic trough collectors are well-documented. One of the principal works performed on this matter is found in \cite{GUVEN1986535}, where, among the main operation errors, profile imperfections and defects that deviate the reflector shape from its ideal parabolic shape are cited. These are considered deterministic in nature and cause a displacement of the focus, contributing negatively to the intercept factor; that is, the ratio of the energy intercepted by the receiver tube to the energy reflected by the mirror \cite{opticsParabolic2019}.

We will consider two main errors for this work that are visible when our 3D model is projected across sections perpendicular to the length of the collector. This view corresponds to the diagram shown in Figure \ref{fig:sectionDiagram}. Under \cite{GUVEN1986535}, the error types we will consider can fall into profile imperfections, which is ultimately what we will measure.

To simulate them, we use simple mathematical approximations that work within the framework of our modeled parabolic trough. True modeling of these errors, such as gravitational sag, requires detailed frameworks that include all structural elements to compute the effects of weight,  supporting beams, and so on.

\begin{enumerate}
    \item \emph{Tilt errors}. We assume that the shape of the panel is not impacted, but rather that its orientation is misaligned compared to the main frame. When the tilt is significant, it can happen that all rays hitting affected panels will miss the receiver. 
    In this case, we will rotate the panel about its centroid, $c$, for angle $\alpha$ along the $z$ axis.
    \begin{equation*}
    v \;\longmapsto\; c + R_z(\alpha)\,\big(v-c\big),
    \end{equation*}
    \item \emph{Dents}. These defects can be caused by the impact of any falling object, for instance, during harsh weather conditions. In our model, they are sharper and more localized than sagging and tilt errors.
    We will model this in terms of a radius parameter $\beta$. 
    
    The function for the error displacement $\psi$ is written in terms of $r$, the radial distance of any point from the center of the bump. The factor $\kappa$ determines where the dent is truncated.
    \begin{equation*}
    \psi(r) =
    \begin{cases}
            \frac{e^{-r^{2}/2\beta^{2}} - e^{-\kappa^{2}/2}}{1 - e^{-\kappa^{2}/2}} & r \leq \kappa\beta  \\
              0 & r > \kappa\beta
    \end{cases}
    \end{equation*}
\end{enumerate}

Figure \ref{fig:sectionalerrors} shows the defect on the 3D panel, and the accompanying profile sections. As expected, these defects cause the expected parabolic shape to distort, and it is that which we hope to detect.

Besides the injected defects, every panel of every collector carries a small baseline tilt. This is not a fault but an attempt to add installation scatter that a real assembled collector would have. We draw it independently for each panel from $\mathcal{N}(0,0.05^{\circ})$ about each axis in the aperture plane. This translates to about one millimeter of translation compared to the perfect parabola. It also adds a perceived width to what would otherwise be a perfectly thin three dimensional model.

\begin{figure}[!htbp]
    \centering
    \includegraphics[width=0.85\linewidth]{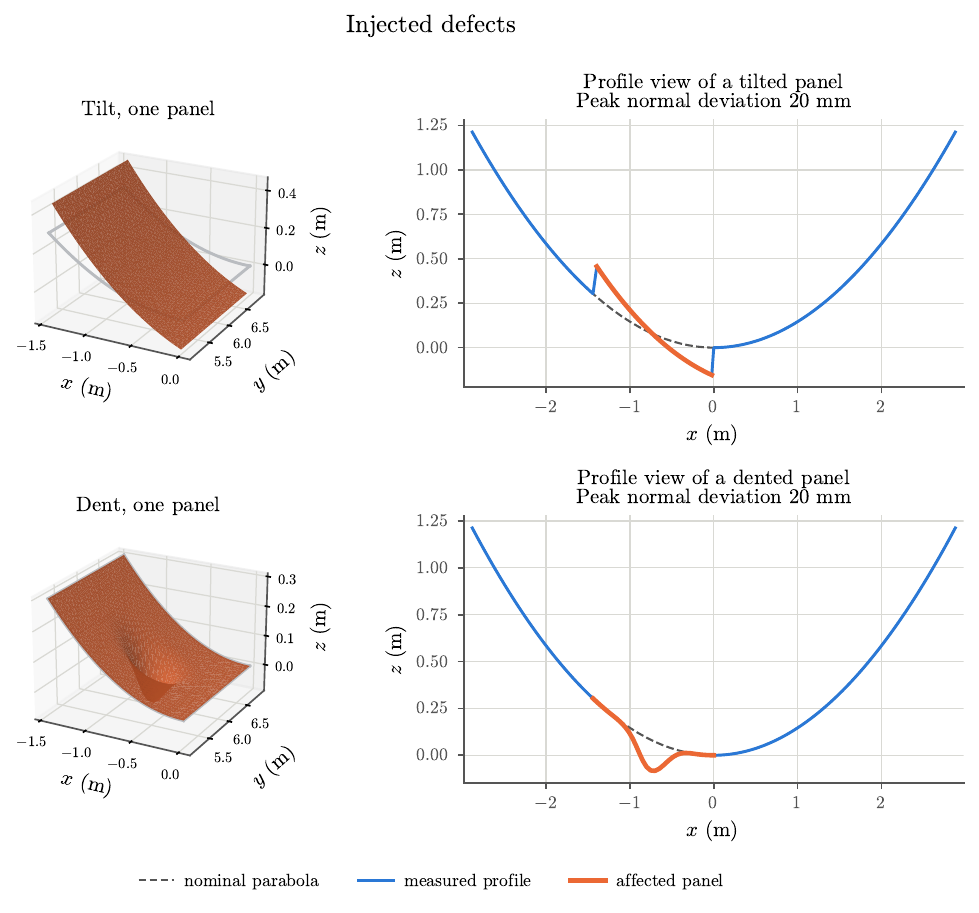}
    \caption{Surface and profile views of the simulated errors, magnified for better visualization.}
    \label{fig:sectionalerrors}
\end{figure}
\newpage
\subsection{Determining threshold values}

In our simulation, we assume the shape of the parabolic trough is ideal, and conditions for experimental setups are factored in with the addition of Gaussian noise, where the variance is in the same order of magnitude as the precision of the device used.

Let $p_i\in\mathbb{R}^3$ denote the ideal position of the $i$-th hit point, and let $d_i\in\mathbb{R}^3$ be the unit vector in the direction of the corresponding laser pulse. The perturbed point is defined as
\begin{equation*}
\bar p_i=p_i+\varepsilon_i d_i,
\end{equation*}
where $\varepsilon_i$ is the measurement error.

We model the errors as independent and identically distributed random variables,
\begin{equation*}
\varepsilon_i\sim\mathcal{N}(0,\sigma^2),
\end{equation*}

where $\sigma$ denotes the standard deviation of the measurement error. Thus, the noise is modeled as a displacement along the laser direction.

We take as null hypothesis that the collector is sound---that its profile is
the nominal parabola, so that the width we measure is due to the measurement
noise alone---against the alternative that the profile departs from it:
\begin{align}
    H_0 &: \text{the reflector profile is the nominal parabola}\\
    H_1 &: \text{the reflector profile departs from it}.
\end{align}
The parabolicity $\rho(\mathcal{P})$ is the statistic by which we decide between the two, $H_0$ being rejected when $\rho(\mathcal{P})>\tau$ for a tolerance $\tau$ fixed in advance. That tolerance is derived below from a simplified noise model and is evaluated against simulated measurements in the section of Results.

We can derive a reference tolerance for a parabolic trough measured by a point cloud containing $N$ points. To obtain a theoretical estimate, we simplify the noise model by neglecting the dependence of the perturbation on the laser direction $d_i$ and the location of each point $p_i$, which would both impact the measured vertical noise. Instead, we assume that the noise affecting each point is characterized solely by an independent scalar error \(\varepsilon_i\), with \(\varepsilon_i\sim\mathcal{N}(0,\sigma^2)\).

Under these assumptions, we consider the range of the $N$ independent errors $R_N$:
\begin{equation*}
R_N=\max_{1\leq i\leq N}\varepsilon_i -\min_{1\leq i\leq N}\varepsilon_i
\end{equation*}
Then, its expected value is given by:
\begin{equation}\label{eq:theoreticalthreshold}
    \mathbb{E}[R_N]=d_2(N)\sigma
\end{equation}
where $d_2(N)$ is the expected range of $N$ independent standard normal random variables, also known as the Tippett factor \cite{tippett1925extreme}. This factor depends on the number of points and grows approximately as $\sqrt{\log N}$ for large $N$.

We will therefore use $d_2(N)\sigma$ as a theoretically motivated reference tolerance. Since the simulated point clouds contain similar numbers of points, we will use a constant tolerance throughout the experiments. As stated in the beginning of this derivation, this value is an approximation derived under a simplified noise model and does not constitute a confidence interval or guarantee a specific false-positive rate.

However, an empirical false-positive rate at this tolerance can be estimated using noise-only datasets, in which no error feature is injected . We define this estimate as:
\begin{equation}\label{eq:alpha}
\widehat{\alpha}
=
\frac{\text{Number of noise-only datasets with }\rho>\tau}
{\text{Total number of noise-only datasets}}
\end{equation}
where $\tau=d_2(N)\sigma$ denotes the chosen reference tolerance and $\rho$ is the computed parabolicity. This amount quantifies how frequently the test identifies a deviation exceeding the tolerance when no feature is present in the simulated data.

\subsection{Algorithmic implementation}

This section describes the four algorithms implemented without delving into the code, which can be found in the repository \url{https://github.com/mmednav/trough_sim}. Algorithms \ref{alg:orientedlp} and \ref{alg:remez} refer to the oriented case; \ref{alg:rotsearch} and \ref{alg:linearised} to the non-oriented one.

\begin{algorithm}[htbp]
\caption{Oriented parabolicity by linear programming}\label{alg:orientedlp}
\begin{algorithmic}[1]
\Require points $P=(x_k,y_k)_{k=1}^{n}$, $n\ge3$
\Ensure the width $\rho^{*}$ and the fitted coefficients $(a,b,c)$

\State $L\gets$ vertices of the convex hull of $P$
  \Comment{candidates for the lower boundary only}
\State $U\gets P$
  \Comment{every point constrains the upper boundary}
\State $u_k\gets(x_k-\bar x)\,/\,\max_j|x_j-\bar x|$ 
  \Comment{for matrix conditioning, the parabolicity remains the same}

\Repeat
  \State solve, in the variables $(a,b,c,t)$,
  \Statex \qquad $\min\;t$
  \Statex \qquad subject to
          $\;a u_k^{2}+bu_k+c\;\le\;y_k$ \quad for $k\in L$,
  \Statex \qquad\phantom{subject to}
          $\;y_k\;\le\;a u_k^{2}+bu_k+c+t$ \quad for $k\in U$,
  \Statex \qquad\phantom{subject to}
          $\;a,b,c$ free, \quad $t\ge0$
  \If{the program is unbounded or not feasible} \Return failure \EndIf
  \If{$a\le0$} $L\gets P$
    \Comment{the hull reduction assumed $a>0$; so it restores $L$}
  \EndIf
\Until{$a>0$ or $L=P$}

\State map $(a,b,c)$ back to the original abscissa

\Return $\rho^{*}=t$ and $(a,b,c)$
\end{algorithmic}
\end{algorithm}

\begin{algorithm}[htbp]
\caption{Oriented parabolicity by Remez exchange}\label{alg:remez}
\begin{algorithmic}[1]
\Require points $(x_k,y_k)_{k=1}^{n}$, $n\ge4$; tolerance $\mathrm{tol}$, maximum number of iterations $M$
\Ensure the width $\rho^{*}$, the parabola's arguments, and its four support points

\State sort the points by abscissa; \textbf{warn} if any two coincide
  \Comment{distinct abscissae are assumed}
\State $u_k\gets(x_k-\bar x)\,/\,\max_j|x_j-\bar x|$
  \Comment{for matrix conditioning, the parabolicity remains the same}
\State $\mathcal{R}\gets$ four indices equally spaced in $1,\dots,n$;\quad
       $\sigma_j\gets(-1)^{j}$

\For{$i=1,\dots,M$}
  \State solve
    $\bigl(u_j^{2}\;\;u_j\;\;1\;\;\sigma_j\bigr)_{j\in\mathcal{R}}
     \,(a,b,c,h)^{\!\top}=(y_j)_{j\in\mathcal{R}}$
    \Comment{$4\times4$; residuals on $\mathcal{R}$ are $\pm h$}
  \State $e_k\gets y_k-(au_k^{2}+bu_k+c)$ for all $k$;\quad
         $k^{*}\gets\arg\max_k|e_k|$
  \If{$|e_{k^{*}}|\le|h|(1+\mathrm{tol})+\mathrm{tol}$} \textbf{break}
    \Comment{no point exceeds the reference level}
  \EndIf
  \State exchange $k^{*}$ into $\mathcal{R}$, preserving the alternation of
         signs
\EndFor

\Return $\rho^{*}=2|h|$, the coefficients mapped back to $x$, and $\mathcal{R}$
\end{algorithmic}
\end{algorithm}
For the non-oriented problems, we present the algorithmic formulation of the two methods proposed.


\begin{algorithm}[htbp]
\caption{Non-oriented parabolicity by rotation search}\label{alg:rotsearch}
\begin{algorithmic}[1]
\Require points $P=(x_k,y_k)_{k=1}^{n}$; bound $\varphi_{\max}$; grid size $N$
\Ensure width $\rho_{\mathrm{rot}}$ and the direction $\varphi$ attaining it

\For{$i=1,\dots,N$}
  \State $P_i\gets\{\,(\,x_k\cos\varphi_i-y_k\sin\varphi_i,\;\;
         x_k\sin\varphi_i+y_k\cos\varphi_i\,)\,\}_{k}$
  \State $\rho_i\gets$ oriented parabolicity of $P_i$, by
         Algorithm \ref{alg:remez}
\EndFor

\State $j\gets\arg\min_i w_i$
\If{$j\in\{1,N\}$} \textbf{warn}: the minimizer lies at the end of the grid
  \Comment{$\varphi_{\max}$ too small} \EndIf
  
\Return $\rho_{\mathrm{rot}}=w_j$ and $\varphi=\varphi_j$
\end{algorithmic}
\end{algorithm}


\begin{algorithm}[htbp]
\caption{Non-oriented parabolicity by linearization}\label{alg:linearised}
\begin{algorithmic}[1]
\Require points $P=(x_k,y_k)_{k=1}^{n}$; nominal parabolic coefficients $a$, $b$;
         trust radius $\varepsilon$; tolerances $C_a$, $C_b$
\Ensure width $\rho_{\mathrm{lin}}$ and the misalignment $\delta$ 

\State $r_k^{0}\gets y_k-a x_k^{2}-b x_k$
  \Comment{residual at the nominal parabola}
\State $u_k\gets x_k+2a\,x_ky_k+b\,y_k$
  \Comment{first-order sensitivity to rotation}

\State solve, in the variables $(\delta,\Delta a,\Delta b,c,t)$,
\Statex \qquad $\min\; t$
\Statex \qquad subject to
        $\;c\;\le\;r_k^{0}+\delta u_k-\Delta a\,x_k^{2}-\Delta b\,x_k\;\le\;c+t$,
        \quad $k=1,\dots,n$,
\Statex \qquad\phantom{subject to}
        $\;|\delta|\le\varepsilon$, \;
        $|\Delta a|\le C_a\varepsilon$, \;
        $|\Delta b|\le C_b\varepsilon$, \;
        $t\ge0$
\If{the program is infeasible or unbounded} \Return failure \EndIf

\Return $\rho_{\mathrm{lin}}=t$ and $\delta$
\end{algorithmic}
\end{algorithm}

\section{Results}\label{sec:results}

In this section, we present the results of our simulations. We create a dataset of 310 parabolic troughs, with an error injected in a selected portion of the troughs. Firstly, we compare the performance of both algorithms for the oriented parabolicity, and then apply the non-oriented heuristics under their respective conditions for the axis orientation. Finally, we propose an experimental threshold under which the troughs can be classified as defective and non-defective.

The characteristics of the dataset used are presented in the following table:

\begin{table}[htbp]
\centering
\begin{tabular}{|l|l|r|r|}
\hline
Error type & Amplitude (mm) & Replicates & Collectors \\
\hline
None & --- & --- & 150 \\
Tilt & 8, 12, 14, 16, 18, 21, 25, 32 & 10 & 80 \\
Dent & 8, 12, 14, 16, 18, 21, 25, 32 & 10 & 80 \\
\hline
\textbf{Total} & & & \textbf{310} \\
\hline
\end{tabular}
\caption{Characteristics of the simulated data set, showing replicates per amplitude size.}
\label{tab:dataset}
\end{table}

To compare both errors, we define their respective amplitudes as the peak departure of the panel surface from its nominal position. For a dent this will the depth at the center of the depression, whereas for a tilt it will be the displacement of the furthest point of the panel. These amplitudes serve as an approximate reference, while the simulation will vary them slightly to avoid making identical collectors.

Note that since our data is a simulation, we know the original axis of symmetry of the ideal parabola. We will call this the \textit{nominal} axis. To compare the performance of the algorithms, we will rotate our system by a random angle $\delta$ and recalculate the data to sit on our randomly misaligned axis. To distinguish between the different algorithms, we will use the notation in Table \ref{tab:rhoNotation}:
\begin{table}[htbp]
\centering
\begin{tabular}{|c|l|l|}
\hline
 & Misalignment present & Computed by \\
\hline
$\rho_{0}$ & No & Algorithm \ref{alg:remez} \\
$\rho_{\delta}$ & Yes, by $\delta$ & Algorithm \ref{alg:remez} over the misaligned data \\
$\rho_{\mathrm{rot}}$ & Yes, by $\delta$ & Algorithm \ref{alg:rotsearch} \\
$\rho_{\mathrm{lin}}$ & Yes, by $\delta$ & Algorithm \ref{alg:linearised} \\
\hline
\end{tabular}
\caption{The four widths compared in our results.}
\label{tab:rhoNotation}
\end{table}

We justify why $\rho_{\delta}$ and $\rho_0$ are computed by Algorithm \ref{alg:remez} instead of \ref{alg:orientedlp} in the following subsection.

\subsection{Performance of LP solver vs Remez}

For our LP, we selected the HiGHS method \cite{huangfu2018}, implemented with SciPy's linear programming solver. As stated in the previous section, we compare 310 collectors, generating a point cloud per trough averaging $83\,386$ points.

Throughout this subsection we assume the orientation is known. We compare the cost of the oriented problem for both algorithms \ref{alg:orientedlp} and \ref{alg:remez}.

\begin{table}[htbp]
\centering
\begin{tabular}{|l|r|}
\hline
Largest relative difference & $2.5\times10^{-14}$ \\
Mean relative difference & $3.8\times10^{-15}$ \\
\hline
Linear program & $0.448$ s, 15 simplex iterations \\
Remez's algorithm & $0.0149$ s, 7 exchanges \\
\hline
\end{tabular}
\caption{The two methods on the point clouds of every collector in the
experiment.}
\label{tab:remezAgreement}
\end{table}

As shown in Table \ref{tab:remezAgreement}, the exchange algorithm ends up being roughly thirty times faster than the linear program solved with HiGHS, the two agreeing on the parabolicity to a relative difference of at most $2.5\times10^{-14}$.

\begin{table}[htbp]
\centering
\begin{tabular}{|l|r|r|r|r|r|}
\hline
 & $\rho$ (mm) & $\max|\Delta|/\rho$ & n. iterations (Rz) & $t_{\mathrm{LP}}$ (s) & $t_{\mathrm{Rz}}$ (s) \\
\hline
\multicolumn{6}{|l|}{\emph{noise varied, $n=83\,400$ points}} \\
\hline
$\sigma=0.5$ mm & 4.32 & $3.6\times10^{-14}$ & 7.2 & 0.395 & 0.011\\
$\sigma=1$ mm & 8.91 & $1.8\times10^{-10}$ & 8.0 & 0.376 & 0.013  \\
$\sigma=2$ mm & 17.05 & $2.5\times10^{-13}$ & 8.2 & 0.375 & 0.012 \\
$\sigma=3$ mm & 26.00 & $1.6\times10^{-10}$ & 8.0 & 0.374 & 0.015  \\
$\sigma=5$ mm & 42.42 & $1.2\times10^{-14}$ & 6.8 & 0.400 & 0.012 \\
$\sigma=10$ mm & 86.43 & $1.7\times10^{-15}$ & 8.0 & 0.397 & 0.011 \\
\hline
\multicolumn{6}{|l|}{\emph{cloud size varied, $\sigma=3$ mm}} \\
\hline
$n=5\,000$ & 21.24 & $6.6\times10^{-14}$ & 6.8 & 0.021 & 0.003 \\
$n=25\,000$ & 22.53 & $3.0\times10^{-10}$ & 8.2 & 0.086 & 0.005 \\
$n=75\,000$ & 25.01 & $2.7\times10^{-14}$ & 6.5 & 0.358 & 0.010 \\
$n=150\,000$ & 26.50 & $7.6\times10^{-15}$ & 9.0 & 0.753 & 0.029 \\
\hline
\end{tabular}
\caption{Comparison of both methods over noise and cloud size variations, showing average times for both.}
\label{tab:remezCost}
\end{table}
Table \ref{tab:remezCost} repeats the comparison across the magnitude of the noise and the size of the cloud. The picture does not change. The agreement holds to between $10^{-10}$ and $10^{-15}$ throughout. It is observed that only the cost of the linear program grows appreciably with the number of points, while the exchange algorithm's number of iterations grows slowly. These results are the reason why we select the exchange algorithm to be the base line of our rotation search, as defined in \ref{alg:rotsearch}.

\subsection{Non-oriented algorithms}

First, we show the results of our rotation search (Algorithm~\ref{alg:rotsearch}). In Figure \ref{fig:rotplotresult}, we display the results over three random collectors. We can observe a clear minimum value, as well as linear behavior on both sides of it.

\begin{figure}[!htbp]
    \centering
    \includegraphics[width=0.7\linewidth]{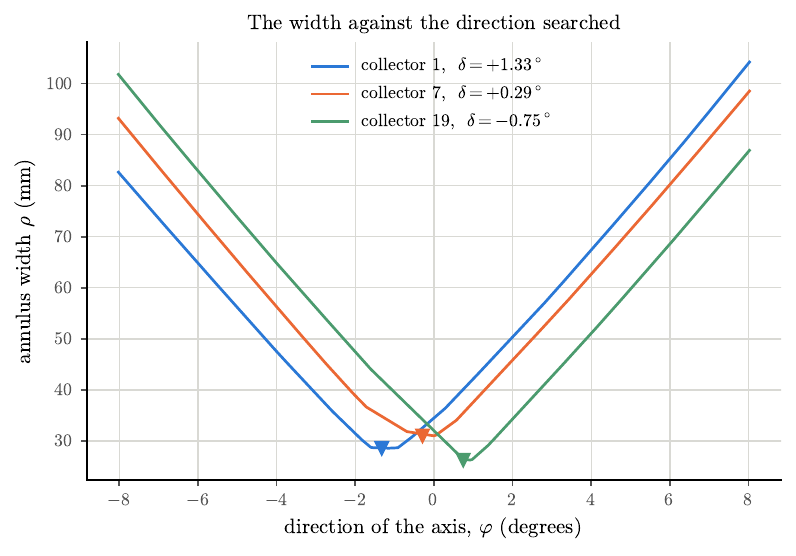}
    \caption{Results of parabolicity over angles, and the minimum achieved for three selected collectors. The zero sits at the nominal axis of symmetry.}
    \label{fig:rotplotresult}
\end{figure}



Figure \ref{fig:plotrecovery} shows how the rotation search performs consistently for non-defective troughs, despite the variation in angle injections of the nominal frame of reference. As expected, the values for $\rho_{\delta}$, which compute the parabolicity over a fixed axis of symmetry, grow with $\delta$, whereas $\rho_{\text{rot}}$ shows a consistent performance. 

\begin{figure}[!htbp]
    \centering
    \includegraphics[width=0.7\linewidth]{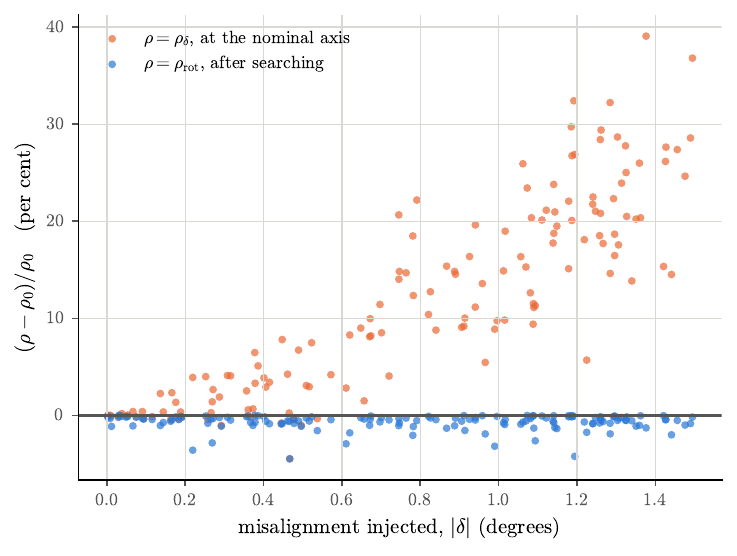}
    \caption{Results of the rotation search against the injected misalignment for the 150 defect-free collectors.}
    \label{fig:plotrecovery}
\end{figure}
\newpage

Figure \ref{fig:comparison} compares the performance of both non-oriented algorithms against each other. In the right-hand plot, we observe that the linearization algorithm is sensitive to tilt errors, and returns larger measures of parabolicity than our rotation search. This difference becomes more evident at larger widths.

\begin{figure}[!htbp]
    \centering
    \includegraphics[width=1\linewidth]{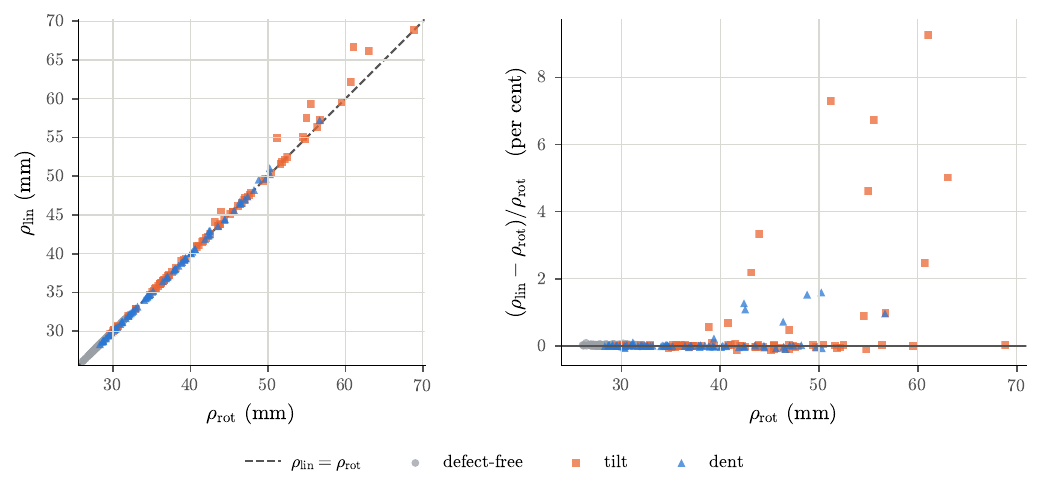}
    \caption{Left: Plot of parabolicity measures returned by both algorithms per collector. Data points are tagged according to Table \ref{tab:dataset}. Right: Deviation of the parabolicity given by the linearization with respect to the rotation search.}
    \label{fig:comparison}
\end{figure}

Lastly, Table \ref{tab:nominalGap} shows the average width for our two non-oriented algorithms and the oriented parabolicity at the nominal axis. Noticeably, our non-oriented approximations did not return a width above $\rho_{0}$ in any of the $310$ collectors, save for eight of them in which the linearization exceeded it by no more than $0.01$ mm.

The direction of this inequality is expected. The non-oriented parabolicity computes a minimum over directions, and the nominal axis is one of them, so the continuous optimum should not exceed $\rho_{0}$.

\begin{table}[htbp]
\centering
\begin{tabular}{|l|r|r|r|r|r|}
\hline
 & $\rho_{0}$ & $\rho_{\delta}$ & $\rho_{\mathrm{rot}}$ & $\rho_{\mathrm{lin}}$
 & $\rho_{0}-\rho_{\mathrm{rot}}$ \\
\hline
No defect ($150$) & $29.10$ & $32.48$ & $28.91$ & $28.91$ & $0.191$ \\
Dent ($80$)       & $37.82$ & $39.51$ & $37.09$ & $37.13$ & $0.731$ \\
Tilt ($80$)       & $45.44$ & $46.58$ & $43.05$ & $43.35$ & $2.384$ \\
\hline
All ($310$)       & $35.57$ & $37.93$ & $34.67$ & $34.76$ & $0.896$ \\
\hline
\end{tabular}
\caption{Mean widths in millimeters, by the defect injected, in the notation of
Table~\ref{tab:rhoNotation}. The last column is the margin by which the
rotation search improves on the nominal axis.}
\label{tab:nominalGap}
\end{table}

This table also shows that the margin is largest for tilted collectors. A tilt is in part indistinguishable from a misalignment of the whole collector, so a search that is free to choose the direction can absorb part of the defect, which is not possible when the axis is fixed at the nominal one.

\subsection{Threshold evaluation}

In this section we turn the parabolicity into a test, by fixing a tolerance
$\tau$ and declaring a collector defective when $\rho>\tau$. We first
compare the theoretical tolerance of \eqref{eq:theoreticalthreshold} against
the one our own data suggests. We then measure the sensitivity of the
resulting test across the amplitudes of the error, and finally present the ROC curve and the confusion matrices.

Throughout, we take as null hypothesis that the collector carries no
injected error, so that the only departures from the ideal parabolic
cylinder are the measurement noise and the small baseline tilts of the
panels. The $150$ collectors of Table \ref{tab:dataset} with no error form
our sample from this hypothesis.

\subsubsection{The parabolicity of a defect-free collector}

The parabolicity of a defect-free collector is not zero. The annulus must
contain every measured point, so its width is essentially the range of the
measurement noise. The first question is whether the theoretical tolerance we derived 
predicts it. Over our $150$ sound collectors we measure $\rho_{0}$ with a
mean of $29.10$ mm and a standard deviation of $1.42$ mm, no collector
falling below $26.10$ mm or above $32.80$ mm. The tolerance of
\eqref{eq:theoreticalthreshold}, evaluated at the median cloud size of
$83\,378$ points and $\sigma=3$ mm, gives $d_{2}(N)=8.72$ and therefore
$\tau=26.17$ mm. The comparison is shown in Figure \ref{fig:null}.

\begin{figure}[!htbp]
    \centering
    \includegraphics[width=0.7\linewidth]{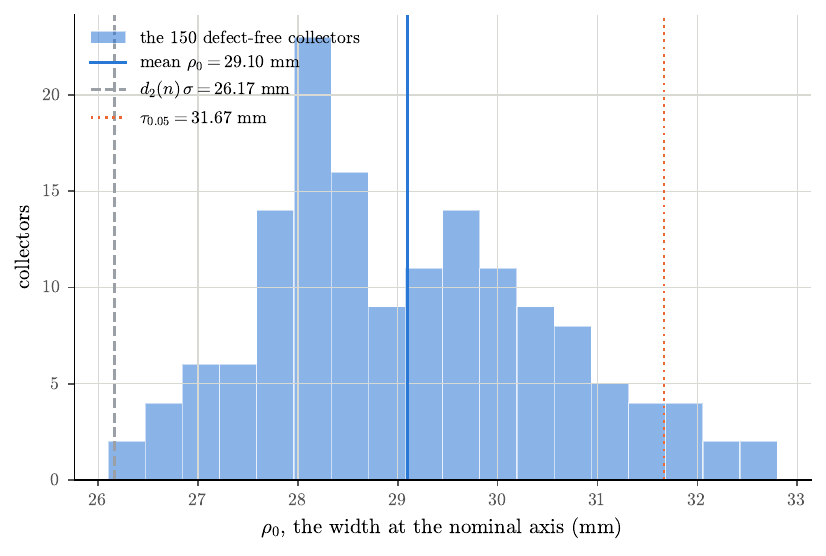}
    \caption{The parabolicity of the $150$ defect-free collectors, with the
    mean of the distribution, the theoretical tolerance $d_{2}(N)\sigma$ and
    the empirical tolerance $\tau_{0.05}$.}
    \label{fig:null}
\end{figure}

The mean of the null distribution thus exceeds the theoretical tolerance, and in fact every collector in it does so but one. Evaluating the empirical false-positive rate \eqref{eq:alpha} at this tolerance gives $\widehat{\alpha}=149/150=0.993$, which means the test rejects almost every non-defective collector. As a decision rule it is unusable, which is the reason we calibrate $\tau$ from the data instead.

The discrepancy shows that the assumptions used to derive \eqref{eq:theoreticalthreshold} do not hold for our residuals. The identity $\mathbb{E}[R_N]=d_2(N)\sigma$ assumes independent, identically distributed normal errors, whereas our residuals contain two departures from this model.

The first thing we find in our model is that the range noise is projected onto the vertical residual along each ray, making its effective standard deviation dependent on ray direction and local mirror slope. Additionally, the baseline tilts introduce coherent deviations across each panel, rather than independent pointwise errors. Overall, this concludes that the discrepancy is primarily due to the spatial structure of the residuals, not their overall variance. For this reason, we opt for an empirical tolerance instead.

 \subsubsection{An empirical tolerance}

We instead set $\tau$ to the $95$th percentile of the sound collectors, giving a false-positive rate of $5\%$ on the null sample. This gives a threshold with the value:
\begin{equation}\label{eq:tauEmpirical}
    \tau_{0.05} = 31.67 \text{ mm}.
\end{equation}
Table \ref{tab:confusion} compares this number with the theoretical tolerance over the full dataset.

\begin{table}[!htbp]
    \centering
    
\definecolor{cmok}{HTML}{2A78D6}
\definecolor{cmerr}{HTML}{EB6834}

\newcommand\CMBox[2][white]{%
  \fcolorbox{black!55}{#1}{%
    \begin{minipage}[c][1.4cm][c]{1.4cm}%
      \centering #2%
    \end{minipage}}%
}

\renewcommand\arraystretch{1}
\setlength\tabcolsep{0pt}
\setlength\fboxsep{0pt}

{\footnotesize
\begin{minipage}[t]{0.48\textwidth}
\centering
\textbf{Theoretical}, $\tau = 26.17$ mm\\[0.6em]
\begin{tabular}{c @{\hspace{0.4em}} c}
        \raisebox{-1.25cm}{\rotatebox{90}{\parbox{2.2cm}{\bfseries\centering actual value}}}
  &
  \begin{tabular}{r @{\hspace{0.5em}}c c @{\hspace{0.5em}}l}
      & \multicolumn{2}{c}{\bfseries predicted} & \\
      & defective & sound & total \\
    defective & \CMBox[cmok!65]{$160$}  & \CMBox[cmerr!6]{$0$} & $160$ \\
    sound     & \CMBox[cmerr!60]{$149$} & \CMBox[cmok!6]{$1$}  & $150$ \\
    total     & $309$ & $1$ & $310$
  \end{tabular}
\end{tabular}\\[0.6em]
$\widehat{\alpha} = 0.993$,\quad accuracy $0.519$
\end{minipage}
\hfill
\begin{minipage}[t]{0.48\textwidth}
\centering
\textbf{Empirical}, $\tau_{0.05} = 31.67$ mm\\[0.6em]
\begin{tabular}{c @{\hspace{0.4em}} c}   
\raisebox{-1.25cm}{\rotatebox{90}{\parbox{2.2cm}{\bfseries\centering actual value}}}
  &
  \begin{tabular}{r @{\hspace{0.5em}}c c @{\hspace{0.5em}}l}
      & \multicolumn{2}{c}{\bfseries predicted} & \\
      & defective & sound & total \\
    defective & \CMBox[cmok!56]{$136$} & \CMBox[cmerr!16]{$24$} & $160$ \\
    sound     & \CMBox[cmerr!9]{$8$}   & \CMBox[cmok!58]{$142$} & $150$ \\
    total     & $144$ & $166$ & $310$
  \end{tabular}
\end{tabular}\\[0.6em]
$\widehat{\alpha} = 0.053$,\quad accuracy $0.897$
\end{minipage}
}

    \caption{Confusion matrices for the two threshold values over the $310$
    collectors, computed on $\rho_{0}$.}
    \label{tab:confusion}
\end{table}

The theoretical tolerance detects every defect, but only by classifying almost every collector as defective, as we previously observed in Figure \ref{fig:null}. This entails that its apparently perfect detection rate comes at the cost of an unacceptably high false-positive rate. On the other hand, the empirical tolerance we computed provides a better balance, retaining most defects while correctly accepting the majority of sound collectors. Note that the performance at false-positives is already predetermined by our threshold, since we computing the matrix over the same set of collectors.

\subsubsection{Sensitivity to the amplitude of the error}

The missed defects are not distributed evenly across the design. Table \ref{tab:detection} and Figure \ref{fig:detection} show the detection fractions at $\tau_{0.05}$ for each amplitude and mechanism. Amplitudes are expressed both in millimeters and relative to the range noise $\sigma$, allowing comparison across instruments for both types of defects tested.

\begin{table}[htbp]
\centering
\begin{tabular}{|r|r|r|r|}
\hline
Amplitude (mm) & $A/\sigma$ & Tilt & Dent \\
\hline
$8$  & $2.67$  & $40\%$  & $10\%$ \\
$12$ & $4.00$  & $100\%$ & $50\%$ \\
$14$ & $4.67$  & $100\%$ & $80\%$ \\
$16$ & $5.33$  & $100\%$ & $90\%$ \\
$18$ & $6.00$  & $100\%$ & $90\%$ \\
$21$ & $7.00$  & $100\%$ & $100\%$ \\
$25$ & $8.33$  & $100\%$ & $100\%$ \\
$32$ & $10.67$ & $100\%$ & $100\%$ \\
\hline
\end{tabular}
\caption{Fraction of the ten replicates detected at $\tau_{0.05}$, by
amplitude and mechanism.}
\label{tab:detection}
\end{table}

\begin{figure}[!htbp]
    \centering
    \includegraphics[width=0.75\linewidth]{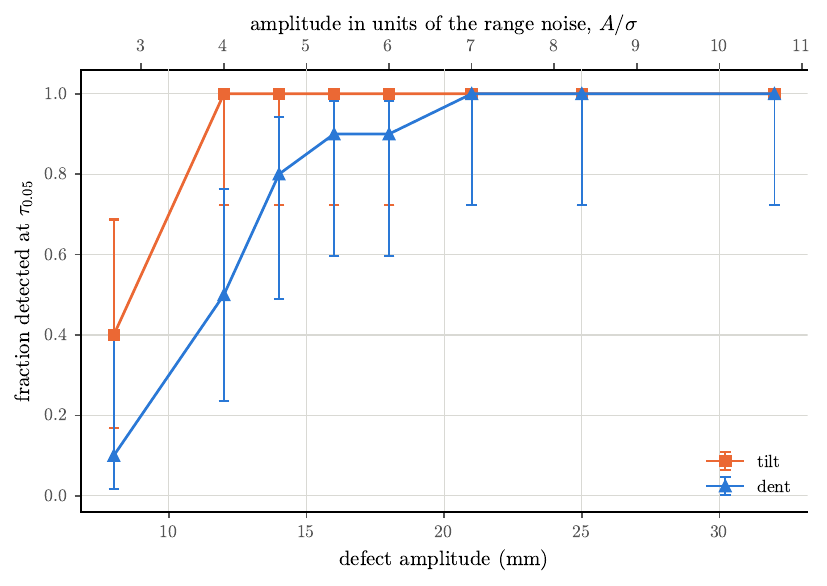}
    \caption{Fraction of collectors detected at $\tau_{0.05}$ against the
    amplitude of the injected error, with Wilson intervals at $95\%$. The
    upper axis gives the amplitude in units of the range noise.}
    \label{fig:detection}
\end{figure}

Both mechanisms become reliably detectable as amplitude increases, but tilts are consistently detected at lower amplitudes than dents. The difference is particularly apparent at intermediate amplitudes, while both mechanisms are detected without exception at the higher end of the design. 

The reason for this is geometric rather than statistical. A tilt rotates an entire panel, producing a displacement that grows across the facet. By contrast, our dent is a truncated Gaussian confined to a disk of fixed radius. Since parabolicity measures the width of an annulus containing the entire surface, detection depends on the largest departure from the best admissible parabola. A mechanism affecting a larger portion of the surface will result in a greater departure for the same nominal amplitude.

\subsubsection{The ROC curve}
Figure \ref{fig:roc} shows the ROC curves for the three estimators, with the operating point of $\tau_{0.05}$ marked. Since the estimators produce different widths, each is evaluated as a score.

\begin{figure}[!htbp]
    \centering
    \includegraphics[width=0.75\linewidth]{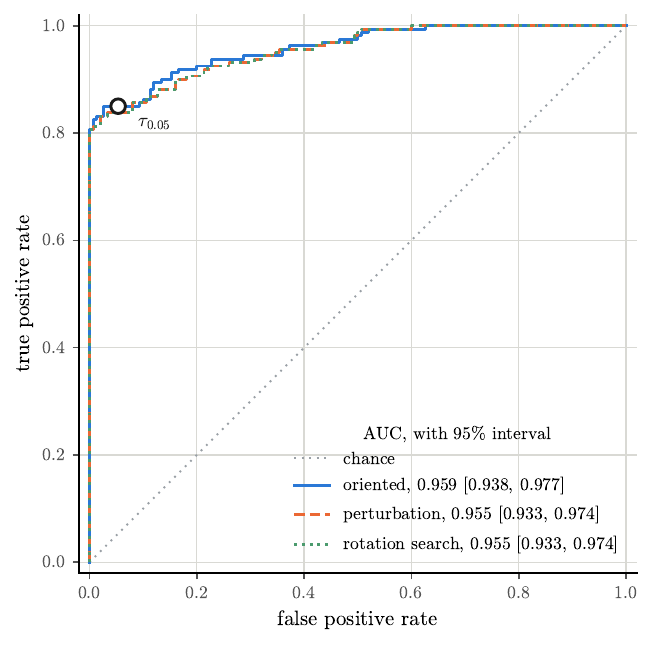}
    \caption{ROC curves for the three estimators, with areas and $95\%$
    bootstrap intervals. The marked point is the operating point of
    $\tau_{0.05}$.}
    \label{fig:roc}
\end{figure}

All three estimators achieve similarly high areas of around $0.96$, indicating that neither the misalignment of the axis of symmetry nor its correction appreciably affects discriminating power. Separating the mechanisms, tilts are more distinguished than dents, consistent with the ordering observed in the detection curves.

These results are conditioned by two key aspects. Firstly, the areas depend on the amplitude distribution of the experimental design, which contains equal numbers of collectors at the smallest and largest amplitudes, so they are comparable across the three estimators but not directly across experiments with different error amplitudes. Secondly, the area measures how well the score ranks defective collectors above sound ones, but it does not determine the reference threshold $\tau$, which is instead set empirically by \eqref{eq:tauEmpirical}.

\subsection{Limitations}

There are multiple limitations to take into account for our experimental setup:
\begin{enumerate}
    \item The simulation engine did not take into account the specularity of the mirror surfaces. Although this can be accounted for in commercial software, obtaining full scans such as the one depicted in \cite{tsl2025}, which shows front-facing 3D scans, through UAV applications may still be limited in favor of terrestrial laser scans due to possible anomalies in the point clouds obtained. In any case, the parabolicity measurements depicted here are still applicable, as both methods generate high-accuracy 3D models of the parabolic trough geometry.
    \item Because the threshold depends directly on the precision of the laser scan, not all defects may be detected. We attempted to reflect this by studying errors of multiple amplitudes.
    \item It is worth noting that we are working with a strictly 2D idealization, whereas solar troughs constitute parabolic cylinders. This entails that certain categories of errors were initially excluded from our case studies, such as axial deformations or warping.
    \item On this line of thought, the 2-dimensional projection of our point cloud relies on certain assumptions about the accuracy of the axis direction. Otherwise, parabolicity can be substantially higher, as any other angle will increase the perceived width of our object. The error associated with this approximation can nonetheless be reduced with finer sectional evaluation, that is, evaluating profiles at fixed widths instead of full troughs.
    
\end{enumerate}

\section{Conclusions and future research}

This work presented a novel geometric framework tailored for shape defect detection in Concentrated Solar Power (CSP) mirror fields. By formalizing the concept of \emph{parabolicity}, we established a formal bridge between computational geometry and optical quality management in solar thermal infrastructure. We formulated this domain-specific alignment challenge into a new class of optimization problems and introduced both exact and approximation algorithms to tackle them efficiently. 


When the direction of the axis is known, the problem is a linear program, and we gave an exchange algorithm that solves it some thirty times faster than a general-purpose solver while agreeing with it to fourteen digits. When the direction is unknown, we showed that uniform angular sampling does not yield a $(1+\varepsilon)$-approximation in general (Theorem~\ref{thm:negative}), and proposed two heuristics, a rotation search and a linearization, which recover the width lost to a misalignment of the frame.

We then evaluated parabolicity as a defect classifier over $310$ simulated collectors. A tolerance predicted from a simple noise model proved unusable, rejecting $149$ of the $150$ defect-free collectors, so we calibrated one from the data instead. It separates defective from sound collectors with an area under the ROC curve of about $0.96$, detecting tilts at lower amplitudes than dents for reasons that are geometric rather than statistical. We therefore showed that parabolicity can be a usable basis for defect detection, provided its tolerance is calibrated on sound collectors.

Several directions for future research emerge from this germinal work:

\emph{Alternative definitions of parabolic width}. The notion of parabolic width adopted in this work, that is, the two parabolic cylinders share a symmetry axis and one is a translate of the other, and the width reduces to the distance along that axis, is only one of several possible measures, and the choice of metric fundamentally shapes the resulting optimization problem. The width admits many definitions depending on the underlying metric: for instance, the metric could be based on the distance between the directrices or on the distance between the parabolas themselves. A systematic study of alternative definitions remains open. For each candidate definition, the key questions are whether the resulting problem admits an efficient exact algorithm, whether its combinatorial structure (e.g., the possible configurations of contact points) can be characterized, and whether approximation schemes with quality guarantees can be derived. Identifying which definitions yield tractable problems, and how the resulting solutions differ in practice, is an open problem of both theoretical and applied interest.

\emph{Algorithmic extensions}. Both the exact and approximation methods proposed here could be improved. In particular, developing approximation schemes with provable quality guarantees for the minimum parabolic annulus problem remains open.

\emph{Higher-dimensional and generalized formulations}. This work has focused on parabolic annuli in the plane. Extending the notion of parabolicity to three-dimensional parabolic cylinders, and to general conic families, would broaden the applicability of the framework to other collector geometries, including heliostats and dish systems.

\emph{Integration with real sensing pipelines}. The synthetic datasets used in our experimental study were generated by simulation. Validating the proposed algorithms on measurement data from operational CSP plants (for example, from drone-based imaging systems) would be an important step toward deployment. In this context, the interaction between measurement uncertainty and the robustness of parabolicity-based defect classification deserves systematic study.

\emph{Learning-based hybrids}. The geometric framework developed here is complementary to the deep-learning approaches that have recently been applied to shape defect detection in CSP plants. Combining the interpretability and provable properties of parabolicity, based methods with the adaptability of learned models, for instance, using parabolicity as a physically grounded feature or as a constraint in a learning pipeline, is a promising direction for real-time, large-scale inspection.

\emph{Broader applications}. Finally, parabolicity as a geometric concept may prove useful beyond CSP, wherever the focus–directrix structure of a surface is physically meaningful. For example, in the design and quality control of optical instruments, antennas, and other reflective systems.

\vspace{1cm}
{\bf Acknowledgements.} This work is supported in part by grants PID2024-159968OB-I00 and TED2021-129182B-I00 funded by MCIN/AEI/10.13039/501100011033 and the European Union NextGenerationEU/PRTR.

\thispagestyle{myplain}
\printbibliography

\end{document}